\documentclass{scrartcl}
\usepackage[top=2cm,bottom=2cm,left=3cm,right=2cm]{geometry}
\usepackage[utf8]{inputenc}
\usepackage[english]{babel}
\usepackage{amsmath}
\usepackage{amsfonts}
\usepackage{amssymb}
\usepackage{graphicx}
\usepackage{import}
\usepackage{pgfplots}
\usepackage{tikz}
\usetikzlibrary{calc,patterns,decorations.pathmorphing,decorations.markings,decorations.pathmorphing,patterns,arrows.meta,spy}
\usepackage[labelformat=simple]{subcaption}

\usepackage{stackengine}
\usepackage[numbers,sort&compress]{natbib}
\usepackage{csquotes}
\usepackage{float} %mehr Kontrolle über Gleitobjekte
\author{Alexander Malik$^1$\thanks{Alexander.Malik@imfd.tu-freiberg.de}, Geralf Hütter$^{1,2}$, Martin Abendroth$^1$, Bjoern Kiefer$^1$\\[1ex]
\small $^1$Institute of Mechanics and Fluid Dynamics, TU Bergakademie Freiberg\\[-0.5ex]
\small Lampadiusstr.~4, 09599 Freiberg, Germany.\\[-0.5ex]
\small $^2$Institute of Civil and Structural Engineering, Brandenburg University of Technology Cottbus-Senftenberg\\[-0.5ex]
\small Konrad-Wachsmann-Allee 2, 03050 Cottbus, Germany.}
\title{Micromorphic FE\textsuperscript{2} Simulation of Plastic Deformations of Foam Structures}
\date{\today}

\def\onedot{$\mathsurround0pt\ldotp$}
\def\cdddot{% three dots 
  \mathbin{\vcenter{\baselineskip.55ex
    \hbox{\onedot}\hbox{\onedot}\hbox{\onedot}%
  }}%
}
\newcommand{\tens}[1]{\boldsymbol{#1}}
\newcommand{\tr}[1]{\mathrm{tr}\left(#1\right)}
\newcommand{\dev}[1]{\mathrm{dev}\left(#1\right)}
\newcommand{\skw}[1]{\mathrm{skw}\left(#1\right)}
\newcommand{\sym}[1]{\mathrm{sym}\left(#1\right)}
\renewcommand{\vec}[1]{\tens{#1}}
\newcommand{\algvec}[1]{\underline{\boldsymbol{#1}}} %algebraic column vector 
\newcommand{\algmatrix}[1]{\underline{\underline{\boldsymbol{#1}}}} %algebraic matrix 
\newcommand{\nvec}[1]{\hat{\algvec{#1}}} %column vector of nodal quantities

\begin{document}
%{\Huge Graphical Abstract}\\[10ex]
%\includegraphics[width=\textwidth]{figures/graphicalAbstract_Micromorphic.pdf}\newpage
\maketitle
%{\footnotesize\tableofcontents}
\begin{abstract}
Capturing and predicting the effective mechanical properties of highly porous cellular media still represents a significant challenge for the research community, due to their complex structural interdependencies and known size effects. Micromorphic theories are often applied in this context to model the inelastic deformation behavior of foam-like structures, in particular to incorporate such size effect into the investigation of structure-property correlations. This raises the problems of formulating appropriate constitutive relations for the numerous non-classical stress measures and determining the corresponding material parameters, which are usually difficult to assess experimentally.

The present contribution therefore alternatively employs a concurrent micromorphic multi-scale approach within the Direct FE\textsuperscript{2} framework to simulate the complex irreversible behavior of foam-like porous solids. The predictions of Cosserat (micropolar) and a fully-micromorphic theory are compared with conventional FE\textsuperscript{2} results and direct numerical simulations (DNS) for complex loading scenarios with elastic, elastic-plastic, and creep deformations. 
\end{abstract}

\section{Introduction}
Computational tools support or even enable the development of application-specific engineered materials and structures and their usage therefore steadily increases. Porous structures, like foams, are widely used in lightweight constructions and as crash absorbers due to their potential to dissipate energy. In these exemplary applications, the mesostructure is in the millimeter range, resulting in a small number of pores in the thinnest dimension. Therefore, a sufficient scale separation is not provided, so that such specimens behave differently than expected from thicker ones. Such so-called size effects are known from numerous experimental studies on foams  \citep{vonHagen1998,Andrews2001,Kirchhof2023,Chen2002, Liebold2016a, Rakow2005, Lakes1983,Lakes1986,Rueger2016,Anderson1994a,Neumann2022,Dillard2006} under different loading conditions.
In order to describe those size-effects at the engineering scales, the conventional Cauchy theory is insufficient but generalized continuum theories are required for this purpose. In particular, the Cosserat theory found wide application \citep{Andrews2001,Lakes1983,Lakes1986,Rueger2016,Anderson1994a,McGregor2014,Diebels2002,Liebenstein2018a} due to its intuitive interpretation. 
 
But even more general micromorphic theories have been used to tackle mentioned effects \citep{Dillard2006,Forest2004,Janicke2013,Huetter2016,Lakes2016, Rokos2018,Huetter2023}, which leads to additional constitutive relations and respective parameters that must be determined.
Even for the linear-elastic modeling, this represents substantial work and it is virtually impossible for inelastic material behavior in the full micromorphic theory. 
Homogenization approaches circumvent this problem but aim to enable predictive numerical simulation based on characteristics of the mesostructure. Such methods have been widely used to determine the effective properties of foams in terms of conventional Cauchy theory, mentioning here only the works by \citet{Gibson1989} and \citet{Warren1997} as most prominent representatives.
In addition, homogenization provides a direct link to the mechanisms at the mesoscale.
Due to its success in the conventional Cauchy theory, numerous studies were devoted to the extension of this approach to generalized continuum theories. Regarding foams and porous materials, these studies are currently limited to elastic behavior \citep{Bouyge2001,Chung2009,Rahali2017,Liebenstein2018a,Rizzi2019,Glaesener2019,SkrzatEremeyev2020,Huetter2023,Huetter2016,Rokos2018} mostly using beam or shell models at the mesoscale \citep{Adomeit1968,Diebels2002,Rahali2017,Liebenstein2018a,Rizzi2019,Glaesener2019} and/or 
Cosserat theory \citep{Diebels2002,Chung2009,Liebenstein2018a,SkrzatEremeyev2020} as the resulting macroscopic continuum (where rotational degrees of freedom are present at mesoscale and macroscale, respectively), with large intersections between these groups.

In contrast, the present contribution addresses the size effects within the irreversible elastic-plastic and creep behavior of foams using a concurrent micromorphic FE\textsuperscript{2} simulation with a resolved unit cell at the mesoscale, including the special case of Cosserat theory for comparison. Thereby, a special focus is given to the predictions of the mechanisms and local fields at the mesoscale. Although the method has already been proposed $25$ years ago \cite{Feyel1999}, the FE\textsuperscript{2} method is still a very active field of research, as well as other methods for concurrent numerical solution of microscale and macroscale, see e.g.~\cite{Feyel1999,Klawonn2019,LANGE2021,Okada2009,Hernandez2014,Lange2024,Wulfinghoff2024}. A review of recent developments can be found in \cite{Raju2021}.

The present article is structured as follows. Section~\ref{sec:Micromorph} starts with a brief review of micromorphic and Cosserat theory including respective homogenization relations. Further, the direct FE\textsuperscript{2} technique is reviewed which is used for the numerical implementation of this theory into a commercial FE code. 
In subsequent Section~\ref{sec:results}, the mesoscopic deformation modes associated with the non-classical micromorphic deformations are investigated for the foam unit cell at the mesoscale and the numerically obtained stiffness values are compared with recent respective analytical estimates from a simplified geometry.
As a next step, the behavior in bending of a beam is considered in the elastic and plastic regime before going to the 
common numerical benchmark problem of a plate with a hole.  
Finally, the model is employed for the complex loading cases of creeping of a flow-through filter and plastic deformations during indentation.
The present contribution is closed by a summary in Section \ref{sec:summary}.

\section{Theory}
\label{sec:Micromorph}
The mesostructure of foams in form of pores and ligaments is the smallest scale under consideration in the present study and is thus termed the \enquote{microscopic} scale in the following. Quantities on this scale are denoted by lowercase letters, whereas capital letters are used for the macroscopic scale. Regarding the micromorphic theory,  the notation of Forest, (cf.~\cite{Forest2002,Forest2003}), is adopted, where possible.

%----------------------------------------------------------
\subsection{Micromorphic theory}\label{sec:fullmicromorph}
%----------------------------------------------------------

In the micromorphic theory of Mindlin \cite{mindlin64} and Eringen \cite{eringen+suhubi64}, 
a second-order tensor field of microdeformation $\tens{\chi}$ is introduced as an additional kinematic DOF besides the classical displacement $\vec{U}$, to achieve a more accurate description of local deformations.
In the case of small deformations, an objective set of strain measures incorporates 
\begin{align}
    \tens{e}&=\tens{H}-\tens{\chi}\ ,&
    \tens{K}&=\tens{\chi}\otimes\vec{\nabla}_{\tens{X}}
    \label{eq:kinematics_fm}
\end{align}
in addition to the conventional strain tensor $\tens{E}\!=\!1/2\,(\tens{H}^T+\tens{H})$ with $\tens{H}\!=\!\vec{U}\otimes\vec{\nabla}_{\tens{X}}$ referring to the displacement gradient. The non-classical measures $\tens{e}$ and $\tens{K}$ can be interpreted as relative deformation and a curvature, respectively.

In the expression for internal virtual power
\begin{align}
\delta W_{\text{int}}&=\overline{\tens{\sigma}}:\delta\tens{E}+\tens{s}:\delta\tens{e}+\tens{M}\cdddot\delta\tens{K}\ ,
\label{eq:power-FM}
\end{align}
the (symmetric) intrinsic stress $\overline{\tens{\sigma}}$, the difference stress $\tens{s}$ and a third-order tensorial hyper stress $\tens{M}$---sometimes also called double stress---occur as respective work-conjugate stress measures.  This expression can equivalently be written in terms of the external stress $\tens{\Sigma}:= \overline{\tens{\sigma}}+\tens{s}^{\mathrm{T}}$ as
\begin{align}
\delta W_{\text{int}}   &=\tens{\Sigma}:\delta\tens{H}-\tens{s}:\delta\tens{\chi}+\tens{M}\cdddot\delta\tens{K}\ ,
\label{eq:power-FM-Sigma}
\end{align}
as favored by \citet{eringen+suhubi64}. From the principle of virtual power \citep{Germain1973}, the balance of linear momentum
\begin{align}
\tens{0}&=\vec{\nabla}_{\tens{X}}\cdot\tens{\Sigma}+\bar{\rho}\,\bar{\tens{f}}\label{eq:linearmomentum}
\end{align}
and a higher-order balance
\begin{align}
\tens{0} &= \vec{\nabla}_{\tens{X}}\cdot\tens{M}+\tens{s}\ ,\label{eq:angularmomentum}
\end{align}
are obtained for the static case, i.e., without inertia, neglecting higher-order body forces here.

If large deformations are considered, respective Green-Lagrange type deformation measures, see \citep{Forest2003}, can be defined as
\begin{align}
    \tens{E}^{\mathrm{GL}}&=\dfrac{1}{2}\left(\tens{H}^T+\tens{H}+\tens{H}^T\cdot\tens{H}\right)\ ,&
    \tens{e}^{\mathrm{GL}}&=\left(\tens{I}+\tens{\chi}\right)^{-1}\cdot\left(\tens{I}+\tens{H}\right)-\tens{I}\ ,&
    \tens{K}^{\mathrm{GL}}&=\left(\tens{I}+\tens{\chi}\right)^{-1}\cdot\left(\tens{\chi}\otimes\vec{\nabla}_{\tens{X}}\right)\ ,
    \label{eq:strainsGreenLagrange}
\end{align}
where the nabla operator $\vec{\nabla}_{\tens{X}}$ refers to the undeformed reference configuration.\footnote{In \citep{Forest2003} and other references, the deformation measures $\tens{\chi}$ and $\tens{e}$ are defined to be of deformation gradient $\tens{F}\!=\!\tens{I}+\tens{H}$ type for large deformations and of displacement gradient $\tens{H}$ type for small deformations. The present contribution uses the latter definition consequently for both cases. The identity  
 $\tens{I}$ is therefore subtracted in  Eq.~\eqref{eq:strainsGreenLagrange}$_2$ from the Cauchy-Green type definition of $\tens{e}$ in \citep{Forest2003}.} 
The second-order identity tensor is denoted as $\tens{I}$. 
In this case,  Eq.~\eqref{eq:power-FM} remains valid for second Piola-Kirchhoff type stresses in conjunction with Green-Lagrange strains \eqref{eq:strainsGreenLagrange}, whereas the balances \eqref{eq:linearmomentum} and \eqref{eq:angularmomentum}  as well as Eq.~\eqref{eq:power-FM-Sigma} are satisfied by their counterpart of first Piola-Kirchhoff type. The conversion between both sets can be found in \ref{sec:stressconversion}.

%-------------------------------------------------
\subsection{Micromorphic homogenization}
\label{sec:micromorphic_homogenization}
%--------------------------------------------
Literally speaking, \emph{homogenization} refers to the replacement of a heterogeneous medium by an equivalent homogeneous one. Following the principle concept of \citet{Hill1963}, this step is achieved by solving a boundary-value problem at the microscale using micro-macro relations. In particular, the theory of \citep{Huetter2017,Huetter2019} is employed here. Therein, the microdeformation tensor at the macroscopic scale
\begin{align}
\tens{\chi}&=\int_{S^\parallel}\vec{u}\otimes\tens{x}\,\mathrm{d}S\cdot\tens{G}^{\chi-1}\label{eq:voidBC-FM}
\end{align}
is introduced as the first moment of the microscopic displacements $\vec{u}$ at the void surface $S^\parallel\!$ . Therein, the tensor of the second geometric moment of the microscopic heterogeneity
\begin{align}
\tens{G}^{\chi}&=\int_{S^\parallel}\tens{ x}\otimes\tens{x}\,\mathrm{d}S
\label{eq:G-fullmicromorph}
\end{align}
ensures that $\tens{\chi}$ has the same units and properties as the macroscopic displacement gradient $\tens{H}$. The latter corresponds to the volume average $\tens{H}\!=\!\langle\vec{u}\otimes\vec{\nabla}_{\vec{x}}\rangle_V$ of its microscopic counterpart making use of the established notation $\langle\bullet\rangle_{\mathcal{V}}:=1/V \int_{\mathcal{V}}\bullet\,\mathrm{d}V$.
The intrinsic stress tensor
\begin{align}
\overline{\tens{\sigma}}&=\langle\tens{\sigma}\rangle_{\mathcal{V}}
\label{eq:sigmabar}
\end{align}
corresponds to the volume average of the microscopic stress field $\tens{\sigma}$ over the unit cell, as in conventional homogenization theory.  In addition, the external stress takes the value
\begin{align}
\tens{\Sigma}&=\dfrac{1}{V} \int_{\partial\mathcal{V}}\tens{x}\otimes\tens{n}\cdot\tens{\sigma}\,\mathrm{d}S\label{eq:sigma-fullmicromorph}
\end{align}
of the first moment of the tractions $\tens{n}\cdot\tens{\sigma}$ at the boundary $\partial\mathcal{V}$ of the unit cell. Therein,  $\tens{n}$ is the outward normal vector at each point $\tens{x}$, which is measured from the geometric center of $\mathcal{V}$. 
Note that $\tens{\Sigma}$ will generally differ from $\overline{\tens{\sigma}}$ by $\tens{s}^T\!=\!\tens{\Sigma}-\overline{\tens{\sigma}}$, which arises by enforcing constraint \eqref{eq:voidBC-FM}.
The same way, the hyper stress
\begin{align}
\tens{M}&=\dfrac{1}{V} \int_{\partial\mathcal{V}}\tens{n}\cdot\tens{\sigma}\otimes\tens{x}\otimes\tens{x}\,\mathrm{d}S
\label{eq:hyperstress}
\end{align}
enters as the second moment of the tractions at the cell boundary. The respective micro-macro relation for the gradient of microdeformation
\begin{align}
\begin{split}
    \tens{K}&=\dfrac{1}{4V}\left(\oint_{\partial\mathcal{V}}\tens{u}\otimes\tens{G}^{-1}\cdot\tens{x}\otimes\tens{n}\,\mathrm{d}S+\oint_{\partial\mathcal{V}}\tens{u}\otimes\tens{n}\otimes\tens{G}^{-1}\cdot\tens{x}\,\mathrm{d}S\right.\\
    &\left.\qquad-\dfrac{2}{2+n}\oint_{\partial\mathcal{V}}\tens{u}\otimes\tens{n}\cdot\tens{x}\,\mathrm{d}S\,\tens{G}^{-1}\right)-\dfrac{1}{2+n}\tens{U}\otimes\tens{G}^{-1}\ ,
    \label{eq:K-micromacro}
\end{split}
\end{align}
utilizes the second geometric moment of the unit cell $\tens{G}=\langle\tens{x}\otimes\tens{x}\rangle_\mathcal{V}$ and depends on the dimension $n$. Note that $\tens{M}$ is symmetric with respect to its last two indices, according to \eqref{eq:hyperstress}. Due to the triple contraction in the virtual power (cf.~Eq.~\eqref{eq:power-FM}) solely the symmetric part of $\tens{K}$ contributes to the internal virtual power. Details are discussed thoroughly in \citep{Huetter2017,Huetter2019,Huetter2023}.

The generalized Hill-Mandel condition (in terms of virtual deformations) then reads
\begin{align}
\langle\tens{\sigma}:\delta\tens{\varepsilon}\rangle_{\mathcal{V}}&=
\delta W_{\text{int}}\ ,
\label{eq:HillMandel-FM}
\end{align}
with the right-hand side according to eqs.~\eqref{eq:power-FM}, \eqref{eq:power-FM-Sigma}, respectively. 
To satisfy this condition, periodic boundary conditions  
\begin{equation}
    \tilde{\vec{u}}\left(\tens{x}^+\right)=\tilde{\vec{u}}\left(\tens{x}^-\right)
    \label{eq:periodicBC}
\end{equation}
for the fluctuation field $\tilde{\vec{u}}$ are are prescribed on opposing homologous points $\tens{x}^+$ and $\tens{x}^-$  of the unit cell boundary $\partial\mathcal{V}$, resulting in the total microscopic displacement field
\begin{align}
\vec{u}(\vec{x})&=\vec{U}+\tens{H}\cdot\vec{x}+\tens{K}:\left(\vec{x}\otimes\vec{x}\right)+\tilde{\vec{u}}(\vec{x})\ .
\label{eq:pBC-FM}
\end{align}
Therein, the total microscopic displacement field relies solely on the symmetric part of $\tens{K}$ in compliance with aforementioned arguments The interested reader is refered to \citep{Huetter2017,Huetter2019,Huetter2023}.

In the large deformation setting, eqs.~\eqref{eq:sigmabar}--\eqref{eq:hyperstress} refer to the stress measures of first Piola-Kirchhoff type.

\subsection{Micropolar (Cosserat) theory}
The micropolar theory is a special case of the full micromorphic theory, for which solely the rotational parts of the microdeformation $\tens{\chi}$ are considered as independent kinematic DOFs. In the small deformation setting, the respective microrotation vector corresponds to the skew-symmetric part $\tens{\Phi}\!=\!-1/2\,\tens{\epsilon}:\tens{\chi}$. An objective set of deformation measures involves the (extended) strain tensor $\tens{E}^{\mathrm{r}}\!=\!\tens{H}^{\mathrm{T}}-\tens{\epsilon}\cdot\vec{\Phi}$  and the second-order curvature tensor $\tens{K}^{\mathrm{r}}\!=\!\vec{\nabla}_{X}\vec{\Phi}$  \cite{schaefer67}, which are work-conjugate in the sense
\begin{align}
    \delta W_{\text{int}}&= \tens{\Sigma}:\delta\tens{E}^{\mathrm{r}}
    +\tens{M}:\delta\tens{K}^{\mathrm{r}}
    \label{eq:powerint-Coss}
\end{align}
to the conventional stress $\tens{\Sigma}$ and to the moment stress $\tens{M}$, respectively.  The balance of  linear momentum~\eqref{eq:linearmomentum} remains valid and the balance of angular momentum becomes
\begin{align}
\tens{0}&=\vec{\nabla}_{\tens{X}}\cdot\tens{M}+\tens{\Sigma}:\tens{\epsilon}\ .
\end{align}
Therein, the symbol $\tens{\epsilon}$ refers to the Levi-Cevita tensor.
For the homogenization step, essentially all relations of Section~\ref{sec:micromorphic_homogenization} remain valid, if double contracted with $\tens{\epsilon}$. In particular, the micro-macro relation for the microrotation follows from \eqref{eq:voidBC-FM} as
\begin{align}
\tens{\Phi}&=\int_{S^\parallel}\tens{x}\times\vec{u}\,\mathrm{d}S\,\cdot\tens{G}^{\chi -1}\ .\label{eq:voidBC-C}
\end{align}
The corresponding kinematic micro-macro relation for the gradient of microrotation follows from the skew part of \eqref{eq:K-micromacro} as discussed thoroughly in \cite{Huetter2019}.
$\tens{K}^{\mathrm{r}}$ enters the periodic boundary condition~\eqref{eq:pBC-FM} as 
\begin{align}
\vec{u}&=\vec{U}+\tens{H}\cdot\vec{x}+\left(\vec{x}\cdot\tens{K}^{\mathrm{r}}\right)\times\vec{x} + \tilde{\vec{u}}\ ,
\label{eq:pBC-C}
\end{align}
where only the deviatoric part is of relevance due to the vector product with $\tens{x}$.
The Hill-Mandel condition~\eqref{eq:HillMandel-FM} remains valid in conjunction with the internal power expression~\eqref{eq:powerint-Coss}. Further details can be found in \citep{Huetter2019}.

If finite deformations are considered, a polar decomposition of Eq.~\eqref{eq:voidBC-FM} is required instead of Eq.~\eqref{eq:voidBC-C}.

\subsection{The Direct FE\textsuperscript{2} method}\label{sec:FE2}
In the FE\textsuperscript{2} approach, the coupled boundary-value problems at the macroscopic and microscopic scales are solved numerically at both scales using finite elements, see review article \citep{Schroeder2014}. 
At the macroscopic scale, the integration $\delta P_{\mathrm{int}}\!=\!\int\delta W_{\mathrm{int}}\,\mathrm{d}V$ for the principle of virtual power $\delta P\!=\!\delta P_{\mathrm{int}}- \delta P_{\mathrm{ext}}\!=\!0$ is performed numerically, i.e.
\begin{align}
    \delta P_{\mathrm{int}}=\sum_{\alpha} w_\alpha J_{\alpha}\delta W_{\mathrm{int},\alpha}
     \label{eq:virtpowerFE}
\end{align}
as the sum over the integration points $\alpha$, each of which are associated with a particular weight $w_{\alpha}$
and value $J_{\alpha}$ of a Jacobian determinant. Therein, $\delta W_{\mathrm{int}}$ is the internal virtual power density, e.g.~$\delta W_{\mathrm{int}}\!=\!\tens{\Sigma}\!:\!\delta\tens{E}$ in Cauchy theory, or including extensions such in eqs.~\eqref{eq:power-FM-Sigma} and \eqref{eq:powerint-Coss} in generalized continuum theories.  
In traditional FE\textsuperscript{2} implementations, a separate Finite Element Analysis (FEA) is performed as a material routine to obtain macroscopic stresses like $\tens{\Sigma}$ (and $\tens{s}$, $\tens{M}$ for the micromorphic theory) for current estimates of deformations $\tens{E}$ (and $\tens{e}$, $\tens{K}$ for the micromorphic theory).  For this purpose, a finite element model of the unit cell is required at each integration point $\alpha$ of the macroscopic elements.  The so-called localization procedure is applied, where the current estimate of the macroscopic deformations is prescribed to the corresponding unit cell using respective boundary conditions, typically periodic ones according to \eqref{eq:periodicBC}. Subsequently, homogenized data (stress and stiffness tensors) are returned to the macroscopic scale. This process is repeated in an iterative way until convergence is achieved at both scales.

An alternative implementation strategy has been proposed recently by Tan and co-workers \citep{tan+raju+lee20} under the name \enquote{Direct FE\textsuperscript{2}}. It inserts the Hill-Mandel condition~\eqref{eq:HillMandel-FM} directly into the macroscopic power expression~\eqref{eq:virtpowerFE}, which yields
\begin{align}
    \delta P_{\mathrm{int}}&=\sum_{\alpha}\dfrac{w_\alpha J_\alpha}{V_\alpha} \, \delta\,p_{\mathrm{int,\alpha}}\ ,
    &\text{with} \qquad 
    \delta p_{\mathrm{int,\alpha}}=\int_{\mathcal{V}_\alpha}\tens{\sigma}:\delta\tens{\varepsilon}\,\mathrm{d}V\ .
    \label{eq:internalWorkinserted}
\end{align}
 In the FE\textsuperscript{2} method, the integral for the total virtual work $\delta\,p_{\mathrm{int,\alpha}}$ of an RVE with domain $\mathcal{V}_\alpha$  in Eq.~\eqref{eq:internalWorkinserted}  is computed numerically after an FE discretization as $\delta\,p_{\mathrm{int,\alpha}}\!=\!\nvec{f}_{\mathrm{int,\alpha}}^{\mathrm{T}}\cdot\delta\nvec{u}_{\alpha}$ through (column) vectors of nodal forces $\nvec{f}_{\mathrm{int,\alpha}}$ and nodal displacements $\nvec{u}_{\alpha}$. 
Consequently, the FE\textsuperscript{2} problem can also be solved by inserting the FE discretized unit cell models directly into the macroscopic model, under the conditions that the necessary kinematic constraints between microscopic nodal displacements $\nvec{u}_{\alpha}$ and macroscopic nodal displacements $\nvec{U}$  are introduced and that the prefactor in \eqref{eq:internalWorkinserted}\textsubscript{1} is ensured to equal unity. In original work \citep{tan+raju+lee20}, the latter condition was satisfied by scaling the unit cell volume $V_\alpha$ to the equivalent volume $w_\alpha J_\alpha$ of the corresponding macroscopic integration point. 
This, however, is not possible with generalized continuum theories, since $V_\alpha$ is an important material parameter itself. At least in plane settings, this problem can be circumvented by adjusting the (nominal) out-of-plane thickness of the microscale elements, while keeping the in-plane dimensions at their physical values, cf.~\citep{Zhi2022}. 

In essence, the direct FE\textsuperscript{2} scheme corresponds to the elimination of the macroscopic stress $\tens{\Sigma}_\alpha$ (as well $\tens{s}_\alpha$ and $\tens{M}_\alpha$) from the governing equations. Consequently, the role of the macroscopic elements degenerates to kinematic constraints between macroscopic nodal displacements (or generally: degrees of freedom) $\nvec{U}$  and their microscopic counterparts  $\nvec{u}_{\alpha}$ to enforce the microscale boundary conditions~\eqref{eq:periodicBC} and potential side-constraints, such as eqs.~\eqref{eq:voidBC-FM} or \eqref{eq:voidBC-C}.
These kinematic constraints remain linear, even in case of large deformations, if two-field deformation measures like $\tens{H}$, $\chi$,  and $\tens{K}$ are used. Objective deformation measures, as defined, e.g., in Eq.~\eqref{eq:strainsGreenLagrange}, are not required in this case, since the unit cell response  $\nvec{f}_{\mathrm{int,\alpha}}$ should itself be invariant to rigid-body motions.
Thus, the big advantage of the direct FE\textsuperscript{2} method is that it can easily be implemented into FE codes, without any changes to the source code and under the usage of all built-in features, just by adding the aforementioned linear constraints in the preprocessing step. A disadvantage is the drastically increased bandwidth of the common system of equations to be solved for $\nvec{U}$ and all $\nvec{u}_{\alpha}$, cf.~\citep{lange+etal22}. Furthermore, the eliminated macroscopic stresses have to be restored in an additional post-processing step, if desired.

The present implementation of the direct FE\textsuperscript{2} framework in the commercial FE software Abaqus is realized via Python scripting. 
On the macroscopic scale, reference DOFs at the integration points $\alpha$ are created for prescribing the set $\algvec{E}_\alpha\!=\![\tens{H}, \tens{\chi},\tens{K}]_\alpha$ of macroscopic deformations to the respective (meshed) unit cell. 
This is realized for the relevant kinematic relations as linear constraints $\algvec{E}_\alpha\!=\!\algmatrix{B}_\alpha \cdot \nvec{U}$ to the respective macro element nodal DOFs $\nvec{U}$. The $\algmatrix{B}_\alpha$ matrix contains the respective values of the shape functions' derivatives according to kinematic relations~\eqref{eq:kinematics_fm}  and, in case of $\tens{\chi}$, the values of the shape functions at the respective integration point themselves, corresponding to the common procedure in FEA.

Regarding the micro-scale, meshed RVEs are included at the position of the macroscopic integration points, and periodic BCs are applied node-wise on their outer boundaries according to Eq.~\eqref{eq:pBC-FM} or \eqref{eq:pBC-C}, and on the inner contour according to Eq.~\eqref{eq:voidBC-FM} or \eqref{eq:voidBC-C}, as linear equations to the previously created reference DOFs $\algvec{E}_\alpha$. 
Details on the discretized version of Eq.~\eqref{eq:voidBC-FM} can be found in \ref{sec:microdeformation_implementation}.

In the work of \citet{tan+raju+lee20}, even the $\algvec{E}_\alpha$ are eliminated by insertion, keeping only $\nvec{U}$ and the nodal DOFs $\nvec{u}_\alpha$ at the microscale. In contrast, we retain the $\algvec{E}_\alpha$ as DOFs in order to extract the macroscopic stresses $[\tens{\Sigma},\tens{s},\tens{M}]$ (of first Piola-Kirchhoff type) as respective work-conjugate \enquote{reaction forces}.
In the planar case under consideration, the set of deformations involves $\algvec{E}_\alpha\!=\![H_{11},H_{22},H_{12},H_{21},\chi_{11},\chi_{22},\chi_{12},\chi_{21},K_{1(12)},K_{2(21)}]$ for the fully micromorphic theory (wherein parentheses around the indices refer to symmetric parts, as usual). Note that other components of $\tens{K}$ do not contribute to the periodic BCs~\eqref{eq:periodicBC} and \eqref{eq:pBC-FM} until further integral constraints are introduced to enforce them, cf.~\citep{Kouznetsova2004,Huetter2019}. For the micropolar theory, the set of deformations comprises $\algvec{E}_\alpha\!=\![H_{11},H_{22},H_{12},H_{21},\Phi_{3},K_{13},K_{23}]$.
In the following examples, quadrilateral elements with quadratic or linear shape functions are used for $\vec{U}$ and $\tens{\chi}$  or $\vec{\Phi}$, along with four Gauss points for numerical integration. 

\section{Results}
\label{sec:results}
Discretely resolving the underlying microstructure of a macroscopic problem is a straightforward approach and yields the most accurate results---at least for known periodic microstructures. Because of the fine meshes usually required in FEA, this approach becomes time-consuming. This notwithstanding, the direct numerical simulation approach is used here to compute reference data of structural behavior for comparison with the micromorphic homogenization results. Computational costs and manageability, however, limit the investigations to the planar case. Since the high-fidelity DNS are considered to represent the benchmark behavior of the examined structures, the considered micromorphic homogenization schemes ought to yield accurate approximations of those results. 

\subsection{Non-classical microscopic deformation modes}
\begin{figure}
    \centering
    \includegraphics[width=0.3\textwidth]{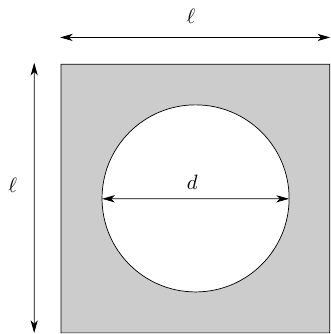}
    \caption{Unit cell to idealize foam microstructure in 2D.}
    \label{fig:RVE}
\end{figure}

\begin{figure}[htp]
    \centering
    \begin{subfigure}{0.49\textwidth}
        \centering
        \includegraphics[width=0.4\textwidth]{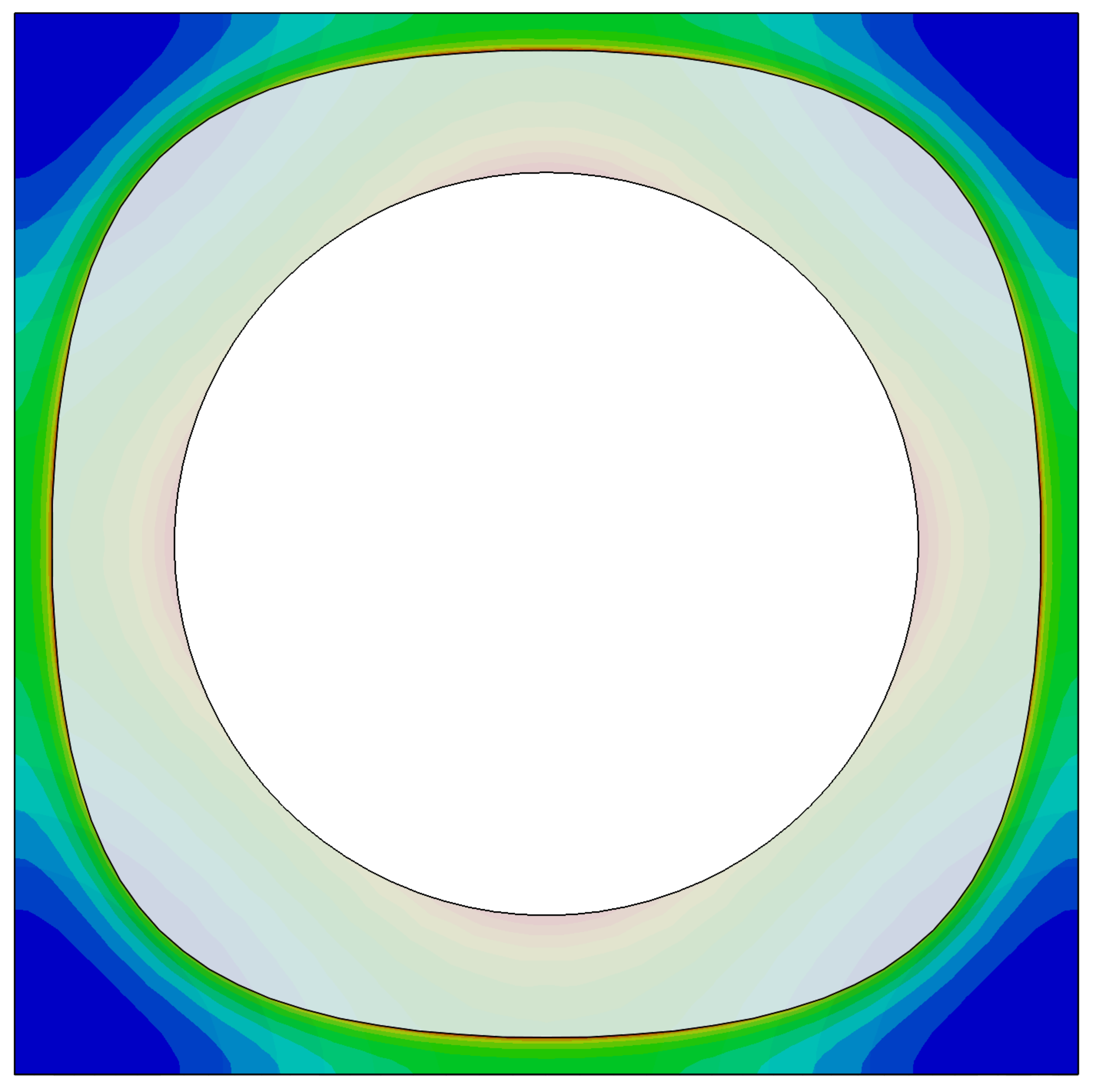}
        \caption{Microdilatation $\chi_{11}\!=\!\chi_{22}$, $\tens{H}\!=\!\tens{0}$}
        \label{fig:dilatation}
    \end{subfigure}
    \begin{subfigure}{0.49\textwidth}
        \centering
        \includegraphics[width=0.4\textwidth]{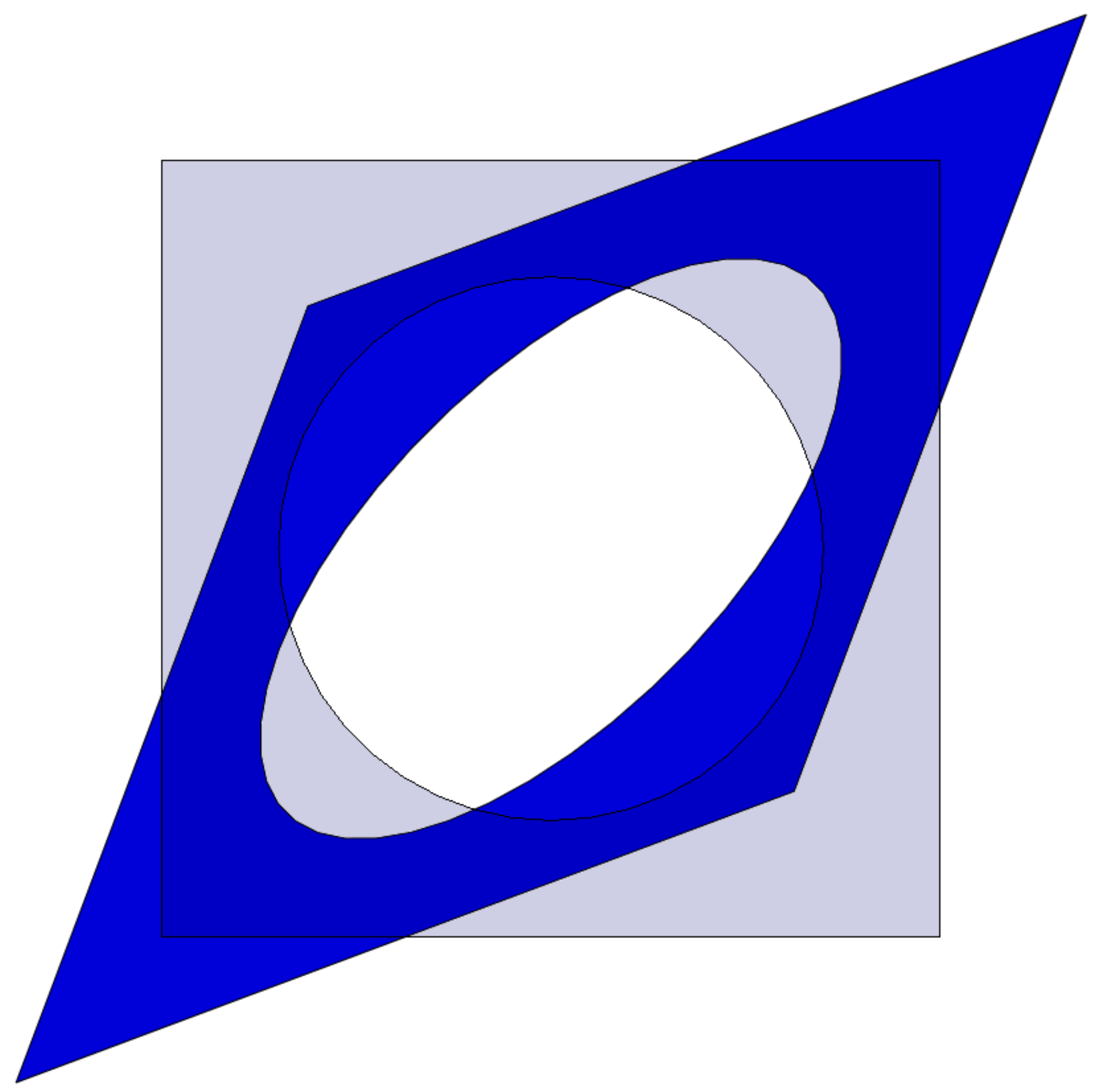}
        \caption{Deviatoric shear $\chi_{12}\!=\!\chi_{21}\!=\!H_{12}\!=\!H_{21}$}
        \label{fig:deviatoricshear}
    \end{subfigure}
    \begin{subfigure}{0.49\textwidth}
        \centering
        \includegraphics[width=0.4\textwidth]{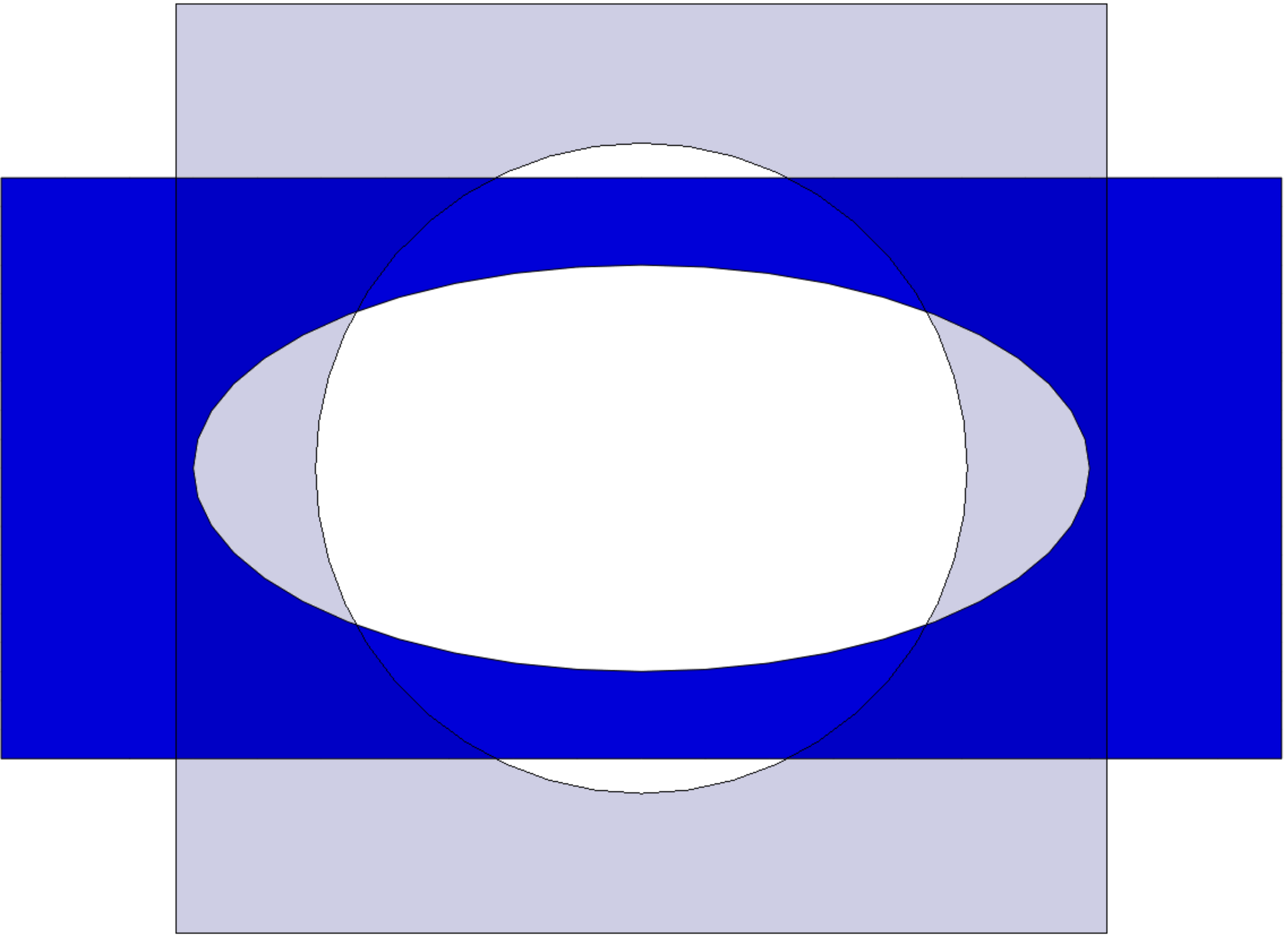}
        \caption{Deviatoric long.~load $\chi_{11}\!=\!-\chi_{22}\!=\!H_{11}\!=\!-H_{22}$}
        \label{fig:deviatoriclongitudinal}
    \end{subfigure}
    \begin{subfigure}{0.49\textwidth}
        \centering
        \includegraphics[width=0.4\textwidth]{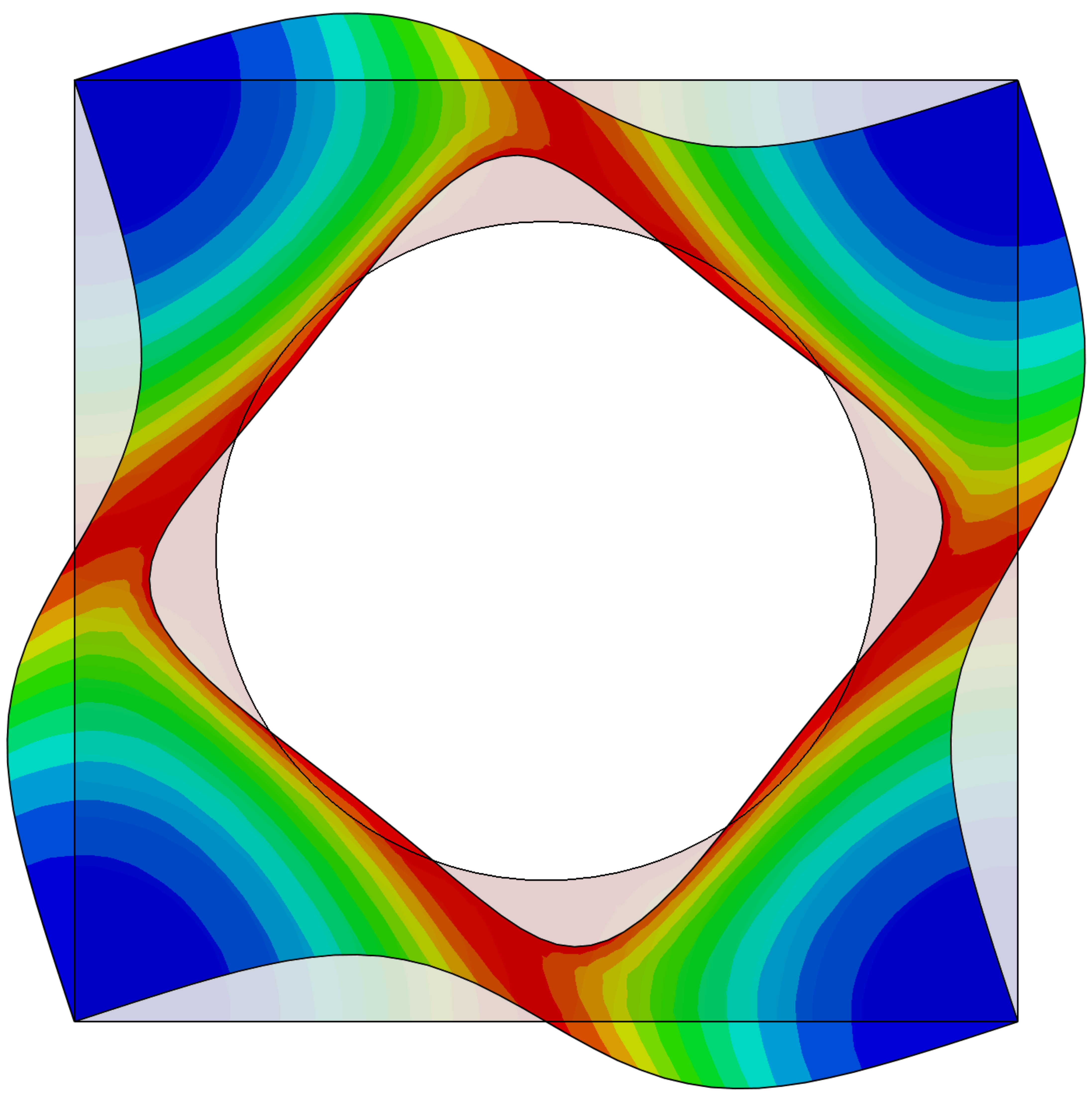}
        \caption{Relative rotation $\chi_{12}\!=\!-\chi_{21}\!=\!\Phi_3$, $\tens{H}\!=\!\tens{0}$}
        \label{fig:relativerotation}
    \end{subfigure}
    \caption{Non-classical deformations of a unit cell (amplified and with a color map representing the von Mises stress in arb. units).}
    \label{fig:RVEdeformation}
\end{figure}

With corresponding extensions, the direct FE\textsuperscript{2} implementation is able to consider micropolar (Cosserat) or micromorphic theory to represent size effects on the macroscale. In order to gain insight into the role of the non-classical deformation measures,
load cases applied on a single unit cell are first investigated, independent of macroscale FEA. The considered pore geometry is shown in Fig.~\ref{fig:RVE} and isotropic, linear-elastic bulk material properties are assumed.   
In this simple case, the microstructure is characterized by the relative porosity $c\!=\!\pi/4\,d^2/\ell^2$, where the unit cell size $\ell$ represents the internal material length scale.
At this point, it is recalled that the displacement gradient $\tens{H}$ controls the outer boundary of the unit cell, \eqref{eq:pBC-FM} and \eqref{eq:pBC-C}, whereas the microdeformation $\tens{\chi}$ controls the deformation of the pore surface (as heterogeneity $S_{\parallel}$ of the unit cell according to eqs.~\eqref{eq:voidBC-FM} and \eqref{eq:voidBC-C})  in an averaged sense. Some exemplary deformations are displayed in Fig.~\ref{fig:RVEdeformation}.

Firstly, Fig.~\ref{fig:dilatation} shows the deformations of the unit cell for a pure microdilatation, where $\chi_{11}\!=\!\chi_{22}$. The outer boundary remains undeformed, due to symmetry, and only the pore expands. The pore does not remain circular, however, since the extension is more pronounced in the diagonal directions, where more deformable bulk material is available.
Figs.~\ref{fig:deviatoricshear} and \ref{fig:deviatoriclongitudinal} show deviatoric loading scenarios with equal values of macrodeformation $\tens{H}$ and microdeformation $\chi$ for each case, corresponding to a vanishing relative deformation $\tens{e}\!=\!0$. Since the square shape of the unit cell introduces cubic anisotropy, two loading scenarios are considered with principal directions along the edges and the diagonals, respectively. Remarkably, the bulk undergoes homogeneous deformations in both cases, in agreement with the more simple unit cell reported in \citep{Huetter2023}.
Finally, Fig.~\ref{fig:relativerotation} shows the effect of a prescribed relative rotation of the pore (in a small displacement setting). Note that the initially straight edges of the unit cell deform into  sinusoidal-like shapes, even though a vanishing macroscopic deformation $\tens{H}\!=\!0$ is prescribed.
Such modes of deformation have been approximated by cubic polynomials related to $\tens{\chi}$ in a number of publications, starting with the seminal work of \citet{Forest1998}. In the present framework, these deformations are predicted as an \emph{outcome} of the micromorphic homogenisation framework, due to the employed combination of periodic boundary conditions and the micro-macro transition relation for $\tens{\chi}$.

In the next step, the quantitative predictions are investigated. For this purpose, the isotropic linear-elastic law is taken as a reference, re-written here in a split into spherical, deviatoric and skew-symmetric parts as
\begin{align}
     \begin{bmatrix}
         \frac{1}{n}\,\tr{\overline{\tens{\sigma}}}\\[0.5ex]  \frac{1}{n}\,\tr{\tens{s}}    \\[0.5ex]
         \dev{\overline{\tens{\sigma}}}\\ \dev{\tens{s}}\\ \skw{\tens{s}}\\ \algvec{M} 
     \end{bmatrix}&=\begin{bmatrix}
         \bar{K}_{E}&\bar{K}_{E\chi}&\textcolor{black!40}{0}&\textcolor{black!40}{0}&\textcolor{black!40}{0}&\textcolor{black!40}{0}\\
        \bar{K}_{E\chi}&\bar{K}_{\chi}&\textcolor{black!40}{0}&\textcolor{black!40}{0}&\textcolor{black!40}{0}&\textcolor{black!40}{0}\\
         \textcolor{black!40}{0}&\textcolor{black!40}{0}&2\,\bar{\mu}&2\,g_2&\textcolor{black!40}{0}&\textcolor{black!40}{0}\\
         \textcolor{black!40}{0}&\textcolor{black!40}{0}&2\,g_2&2\,\bar{\mu}_{\chi}&\textcolor{black!40}{0}&\textcolor{black!40}{0}\\
        \textcolor{black!40}{0}&\textcolor{black!40}{0}&\textcolor{black!40}{0}&\textcolor{black!40}{0}&\kappa&\textcolor{black!40}{0}\\
         \textcolor{black!40}{0}&\textcolor{black!40}{0}&\textcolor{black!40}{0}&\textcolor{black!40}{0}&\textcolor{black!40}{0}&a
     \end{bmatrix}\cdot\begin{bmatrix}
         \tr{\tens{E}}\\ \tr{\tens{e}}\\ \dev{\tens{E}}\\ \dev{\tens{e}}\\ \skw{\tens{e}} \\ \algvec{K}
     \end{bmatrix}
\ ,\end{align}
with the non-classical moduli $\bar{K}_{E}$, $\bar{K}_{\chi}$ and $\bar{K}_{E\chi}$ for the hydrostatic stresses 
$1/n\,\mathrm{tr}(\tens{\bullet})$, 
in dimension $n\!=\!2$, and dilatational strains 
$\mathrm{tr}(\tens{\bullet})$. 
The deviatoric moduli $\bar{\mu}$, $g_2$ and $\bar{\mu}_{\chi}$ are used to relate the (symmetric) deviatoric stresses and strains. Note that the purely skew difference stress tensor $\skw{\tens{s}}$
appears in linear relation to the corresponding strain component, via the Cosserat coupling modulus $\kappa$. 
The same is true for the hyperstress components $\algvec{M}\!=\![M_{1(12)}, M_{2(12)}]$ and the corresponding gradient of microdeformation $\algvec{K}\!=\![K_{1(12)}, K_{2(12)}]$, which are related by the hypermodulus $a$.

\begin{figure}[hp]
    \centering
    \begin{subfigure}{0.49\textwidth}
    \centering
    \includegraphics{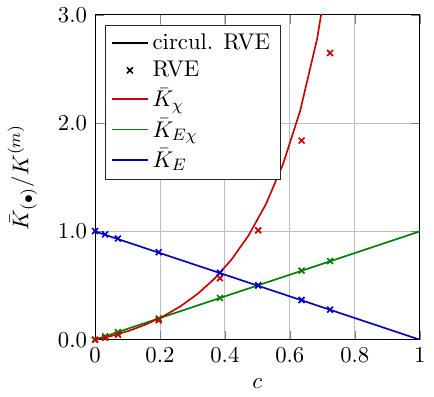}
    \caption{Dilatational moduli}
    \label{fig:DilatationModules}
    \end{subfigure}
    \begin{subfigure}{0.49\textwidth}
    \centering
    \includegraphics{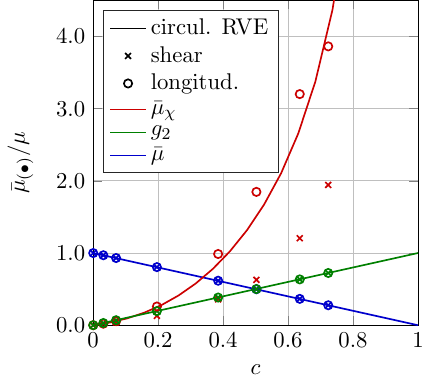}
    \caption{Deviatoric strain moduli}
    \label{fig:DeviatoricModules}
    \end{subfigure}
    \begin{subfigure}{0.49\textwidth}
    \centering
    \includegraphics{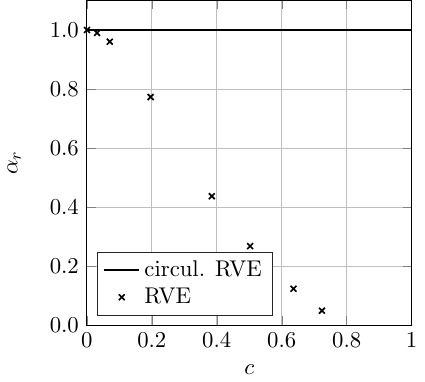}
    \caption{Zener ratio $\alpha_r\!={2\,\bar{\mu}_{\mathrm{eff}}\,(1+\overline{\nu}_{\mathrm{eff}})}/{\bar{E}_{\mathrm{eff}}}$}
    \label{fig:Zenerratio}
    \end{subfigure}\vspace*{5mm}
    \begin{subfigure}{0.49\textwidth}
    \centering
    \includegraphics{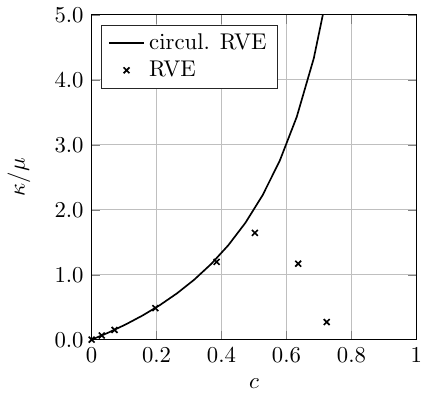}
    \caption{Relative rotation modulus}
    \label{fig:RelativeRotationModules}
    \end{subfigure}
    \begin{subfigure}{0.49\textwidth}
    \centering
    \includegraphics{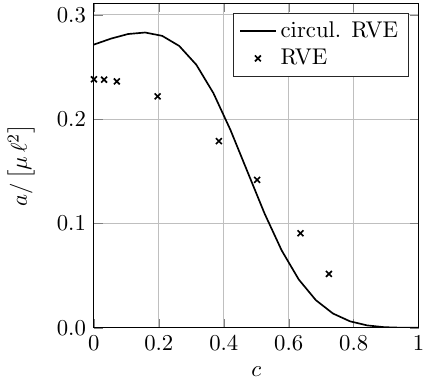}
    \caption{Hyperstress modulus}
    \label{fig:HyperstressModules}
    \end{subfigure}
    \caption{Moduli of an analytically derived circular unit cell and of the considered square unit cell resulting from direct FE\textsuperscript{2} simulations for different porosities.}
    \label{fig:Modules}
\end{figure}

Values of these moduli obtained from unit cell computations are shown in Fig.~\ref{fig:Modules}, for a varying relative porosity $c$.
For comparison, the analytical solutions for a circular unit cell, reported in \cite{Huetter2023}, have also been included in the plots.
As expected, the present solution complies well with the analytical solution for approximately $c\leq 0.2\dots0.4$, depending in detail on the load case and desired accuracy. 
As shown in Figs.~\ref{fig:DilatationModules} and \ref{fig:DeviatoricModules}, the strain moduli $\bar{K}_E$ and $\bar{\mu}$ as well as $\bar{K}_{E\chi}$ and $g_2$ computed for the square cell microstructure even comply exactly with the respective linear relations (corresponding to the Taylor-Voigt bound) of the analytical solution for the circular cell. This finding is consistent with the homogeneous deformation states obtained at the microscale, as previously shown in Figs.~\ref {fig:deviatoricshear} and \ref{fig:deviatoriclongitudinal}, and can thus be generalized to hold independently of the shape of the unit cell. 
Deviations between the analytical solution for a circular unit cell and the present solution can be observed only in the micromoduli $\bar{K}_{\chi}$ and $\bar{\mu}_{\chi}$. For the latter, it is recalled that the present square unit cell only has cubic symmetry, so that different values of $\bar{\mu}_{\chi}$ are obtained depending on whether the principal axes of loading are parallel to the unit cell edges (\enquote{longitud.}) or rotated by 45\textdegree,  as in the shearing (\enquote{shear}) case. 
The anisotropy of a cubic cell can be expressed by the Zener ratio 
$\alpha_r\!=\!{2\,\bar{\mu}_{\mathrm{eff}}\,(1+\overline{\nu}_{\mathrm{eff}})}/{\bar{E}_{\mathrm{eff}}}$, which is analogously adapted to the investigated planar scenario.
It actually corresponds to the stiffness ratio of deviatoric deformation modes along both planes of symmetry in a cubic structure. In terms of the employed Lam\'e parameters, it can thus be written as
\begin{align}
    \alpha_r=\dfrac{\bar{\mu}^{\text{shear}}_{\mathrm{eff}}}{\bar{\mu}^{\text{longitud.}}_{\mathrm{eff}}}\ .
\end{align}
For the micromorphic theory, the effective shear moduli are related to the non-classical moduli by the relation 
$\bar{\mu}_{\mathrm{eff}}\!=\!\bar{\mu}-g_2^2/{\bar{\mu}_{\chi}^{\text{shear}}}$, corresponding to the situation of $\dev{\tens{s}}\!=\!0$.
The values of $\alpha_r$ obtained for the unit cell under investigation are plotted in Fig.~\ref{fig:Zenerratio}. It can be seen that the pore-free unit cell ($c\!=\!0$) behaves isotropically ($\alpha_r\!=\!1$), but the anisotropy increases (corresponding to decreasing Zener ratio values) with increasing porosity $c$.
Apparently, the influence of the thin ligament between pore surface and edge of the chosen unit cell on the effective stiffness is stronger in the shearing case than in longitudinal loading of the unit cell. This effect is even more pronounced regarding the resistance $\kappa$ against relative rotation, as plotted in Fig.~\ref{fig:RelativeRotationModules}. For the non-circular RVE, the values of this Cosserat coupling modulus even reach a maximum at a porosity of $c\approx0.5$, beyond which they drop to zero at the maximally realizable porosity of $c\!=\!\pi/4\!=\!78.5\%$ (the point at which the pore surface intersects the unit cell boundary).

A qualitative, and even relatively close quantitative, agreement between the hyperstress moduli of the present cubic and the reference circular unit cells also found in Fig.~\ref{fig:HyperstressModules}.\footnote{The constraints of the gradient of microdeformation \eqref{eq:K-micromacro} are not enforced explicitly for the quadratic unit cell. Hence, the results are not immediately comparable with the circular unit cell but convertible, cf.~\cite{Huetter2023}.
}
In contrast to the previous plots a difference in moduli is observed between the two unit cell types for vanishing porosity ($c\!=\!0$). 
However, in this case all coupling moduli $\bar{K}_{\chi}$, $\bar{\mu}_{\chi}$, $\bar{K}_{E\chi}$ and $g_2$ vanish so that the macroscopic continuum degenerates to a Cauchy continuum (while keeping resistance to hyperstresses $\tens{M}$ at a boundary due to non-vanishing $a$).
Furthermore, the dependency of the hyperstress moduli $a$ on the size and shape of the employed unit cell is a well known-problem with quadratic boundary conditions \eqref{eq:pBC-FM}. Using the smallest possible unit cell has been established as a pragmatic solution -- and finally Fig.~\ref{fig:HyperstressModules} shows that the differences between the predictions for both shapes are limited.

%----------------------------------------------------------
\subsection{Pure bending of a beam}
\begin{figure}
  \centering
  \begin{subfigure}{0.74\textwidth}
    \centering
    \includegraphics[width=0.9\textwidth]{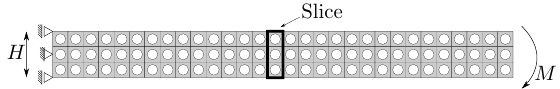}
    %\caption{Direct numerical simulation (DNS) with highlighted slice}
    \caption{}
    \label{fig:DNSbeam}
  \end{subfigure}
%\end{figure}
% \begin{figure}
%     \centering
%     \def\svgwidth{0.15\textwidth}
%     \import{figures/}{DNSbeamslice.pdf_tex}
%     \caption{DNS beam slice.}
%     \label{fig:DNSbeamslice}
% \end{figure}
%\begin{figure}
  \begin{subfigure}{0.24\textwidth}
    \centering
    \includegraphics[width=0.9\textwidth]{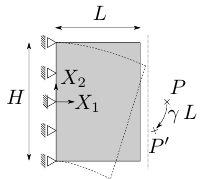}
    %\caption{Boundary conditions for DNS and direct FE\textsuperscript{2}}
    \caption{}
    \label{fig:beamBC}
  \end{subfigure}
  \caption{Pure bending of a beam: \subref{fig:DNSbeam} direct numerical simulation (DNS) with highlighted slice and \subref{fig:beamBC} boundary conditions for DNS and direct FE\textsuperscript{2}.}
  \label{fig:purebending}
\end{figure}

%----------------------------------------------------------
Because of its simplicity, the bending of a beam at small strains has been investigated as a benchmark problem in numerous studies on size effects, and so do we here. In particular, the case of pure bending is investigated as depicted in Fig.~\ref{fig:purebending}, where the macroscopic fields of deformation and stress are independent of the longitudinal direction $X_1$. 
This property is favorable in particular for direct numerical simulations with resolved microstructure, for which a beam slice is sufficient, as indicated in Fig.~\ref{fig:DNSbeam}. In this case, the deformations follow the classical solution of the theory of elasticity \emph{on average}, but with periodic fluctuations.
Hence, the left- and right-hand edges of the slice are constrained periodically as, see \citep{pham+huetter21},
\begin{align}
\begin{split}
U_1^\text{left}(X_2)&=0\ ,\quad
U_2^\text{left}(X_2=0)=0\ ,\\
U_1^\text{right}(X_2)&=-\gamma \,L \,X_2\ ,\quad
U_2^\text{right}(X_2)=U_2^\text{left}(X_2)+\dfrac{1}{2}\, \gamma \,L^2
\end{split}
\label{eq:purebendBCperiodic}
\end{align}
in terms of a prescribed curvature $\gamma$ as shown schematically in Fig.~\ref{fig:beamBC}. A controlling point $P$ is used to apply $\gamma$ in FEA to access the total bending moment $M$ of the investigated beam as the work-conjugate quantity to the mean rotation $\gamma L$ of the right-hand side surface. 
Regarding the continuum models, it is known for the Cosserat continuum \citep{Gauthier1975} and for the micromorphic continuum \citep{Rizzi2021,Huetter2023} that the displacement field for pure bending does not deviate from the classical solution. A  respective numerical solution can thus be obtained by the FE\textsuperscript{2} method using the displacement boundary conditions~\eqref{eq:purebendBCperiodic} together with the additional higher-order periodicity conditions: $\sym{\chi}^\text{left}(X_2)\!=\!\sym{\chi}^\text{right}(X_2)$ and $\vec{\Phi}_3^\text{left}(X_2)\!=\!\vec{\Phi}_3^\text{right}(X_2)+\gamma L$. 
For the fully-micromorphic theory, a boundary layer occurs at the top and bottom macroscopic surfaces of the beam, which needs to be resolved by a sufficient number of elements in the numerical analysis. A mesh convergence study showed that five elements over the height of the beam provided an accuracy of $98.5\%$ for the global $M$ response and this resolution was therefore used in the following studies.

\begin{figure}
        \centering
        \begin{subfigure}{0.15\textwidth}
        \hspace*{0.8em}
        %\abqscaletenminmax{6.79}{-6.79}{$\sigma_{11}/(E_{\mathrm{eff}\gamma H})$}{0.5}{1}
        \includegraphics{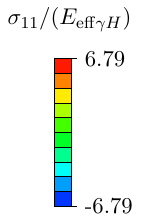}
        \end{subfigure}
        \begin{subfigure}{0.19\textwidth}
        \centering
        \includegraphics[width=\textwidth]{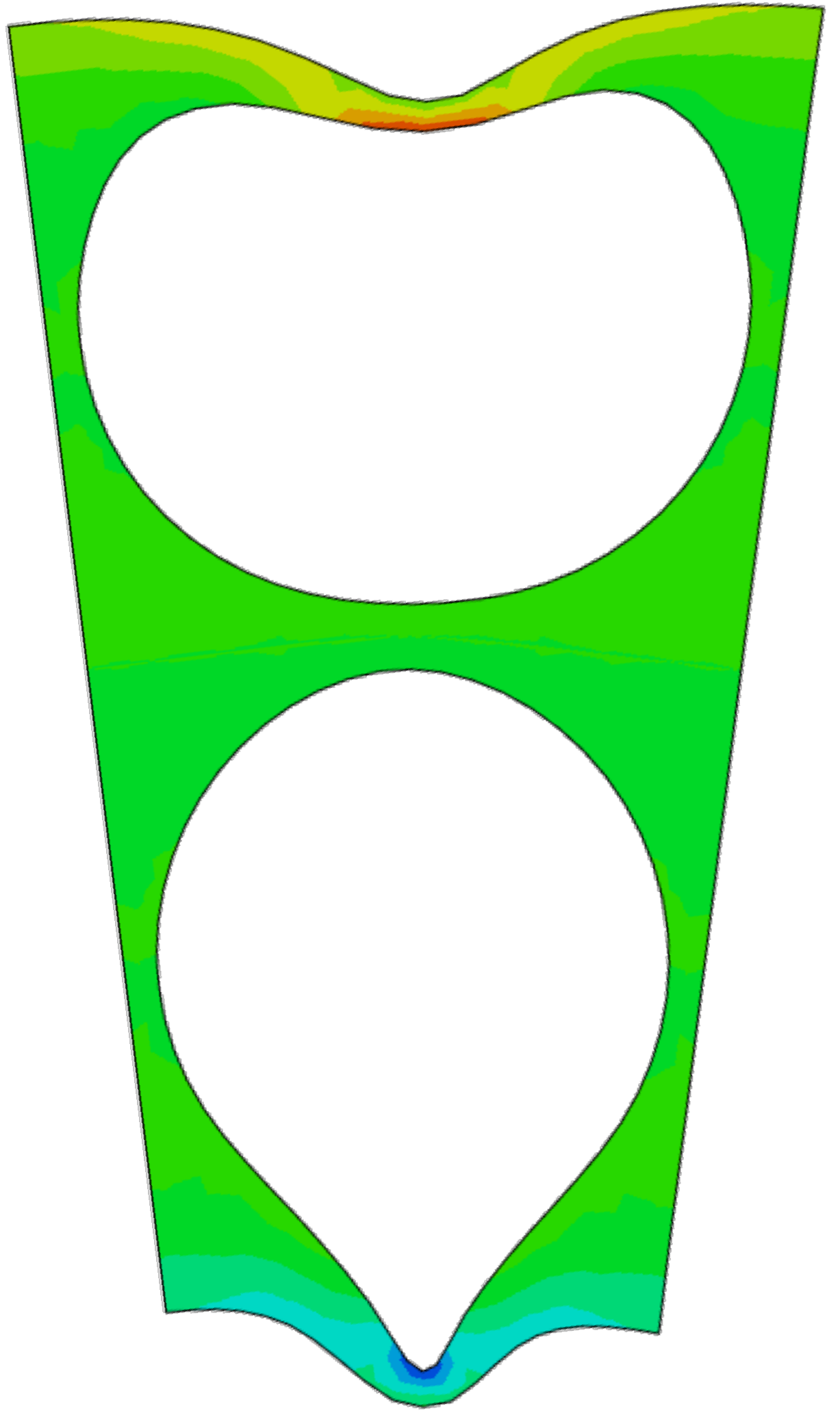}
        \caption{DNS}
        \label{fig:ComparisonDeformation_DNS}
        \end{subfigure}
        \begin{subfigure}{0.19\textwidth}
        \centering
        \includegraphics[width=0.97\textwidth]{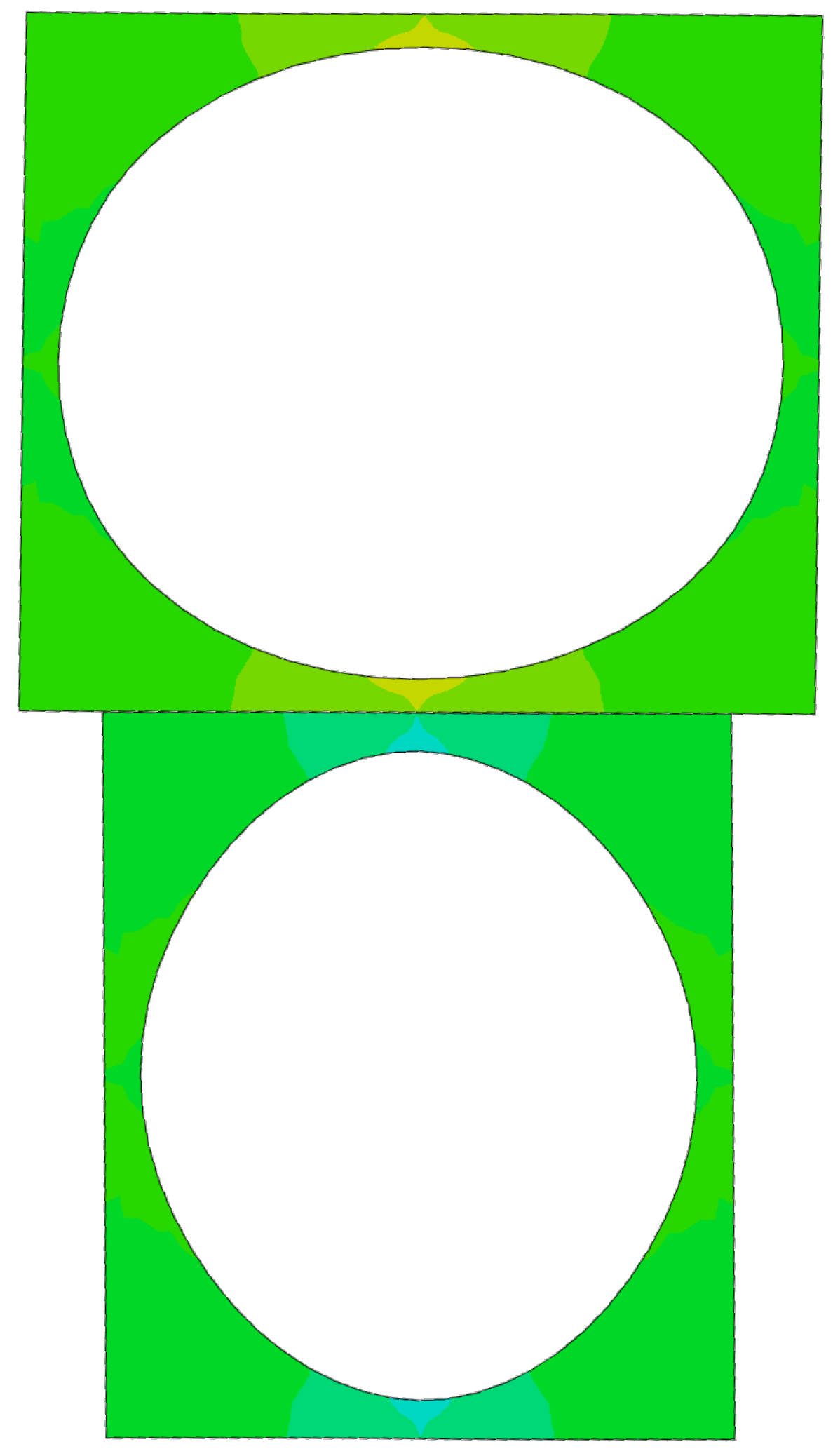}
        \caption{Cauchy}
        \label{fig:ComparisonDeformation_Cauchy}
        \end{subfigure}
        \begin{subfigure}{0.19\textwidth}
        \centering
        \includegraphics[width=1.03\textwidth]{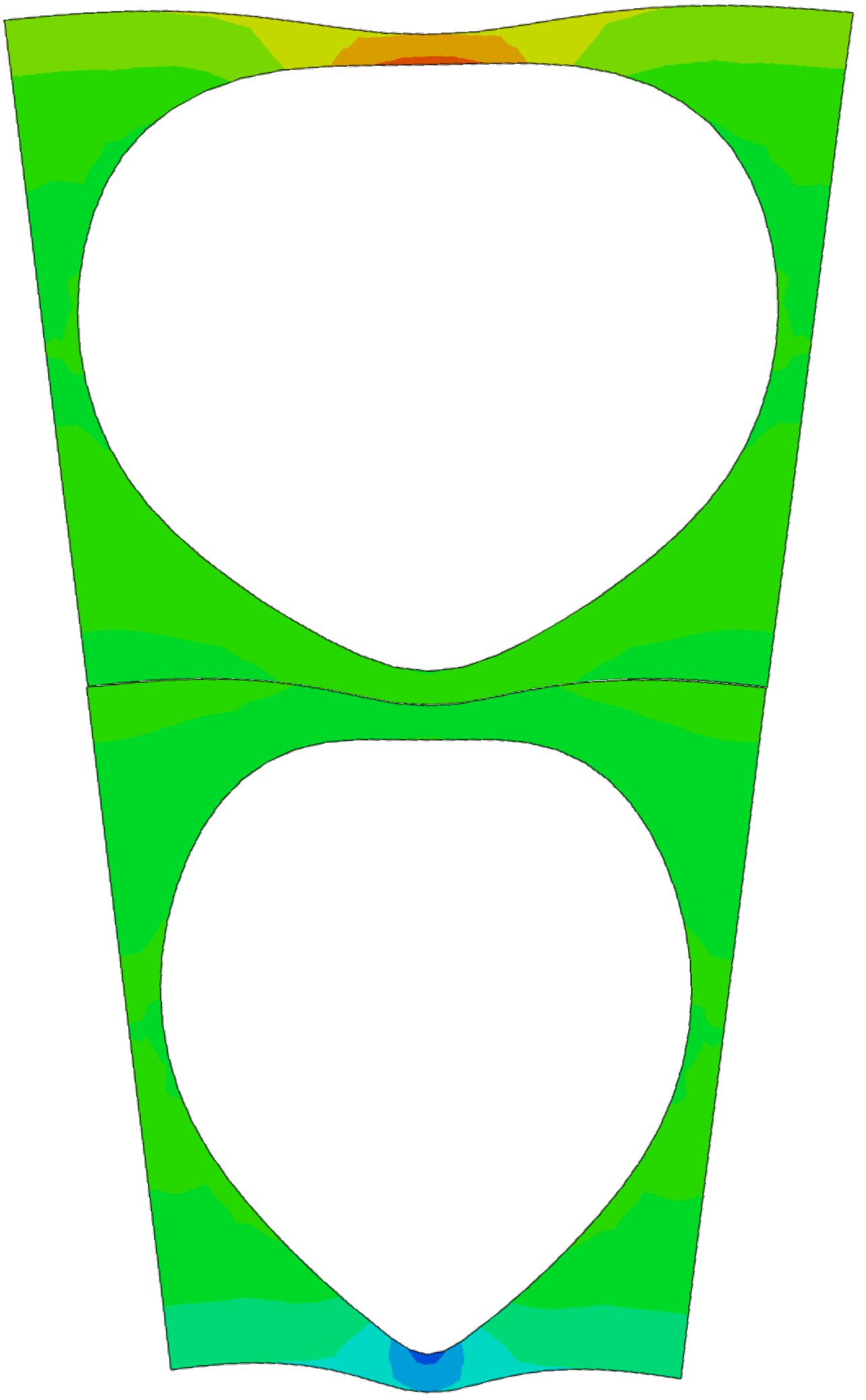}
        \caption{Cosserat}
        \label{fig:ComparisonDeformation_Cosserat}
        \end{subfigure}
        \begin{subfigure}{0.19\textwidth}
        \centering
        \includegraphics[width=1.07\textwidth]{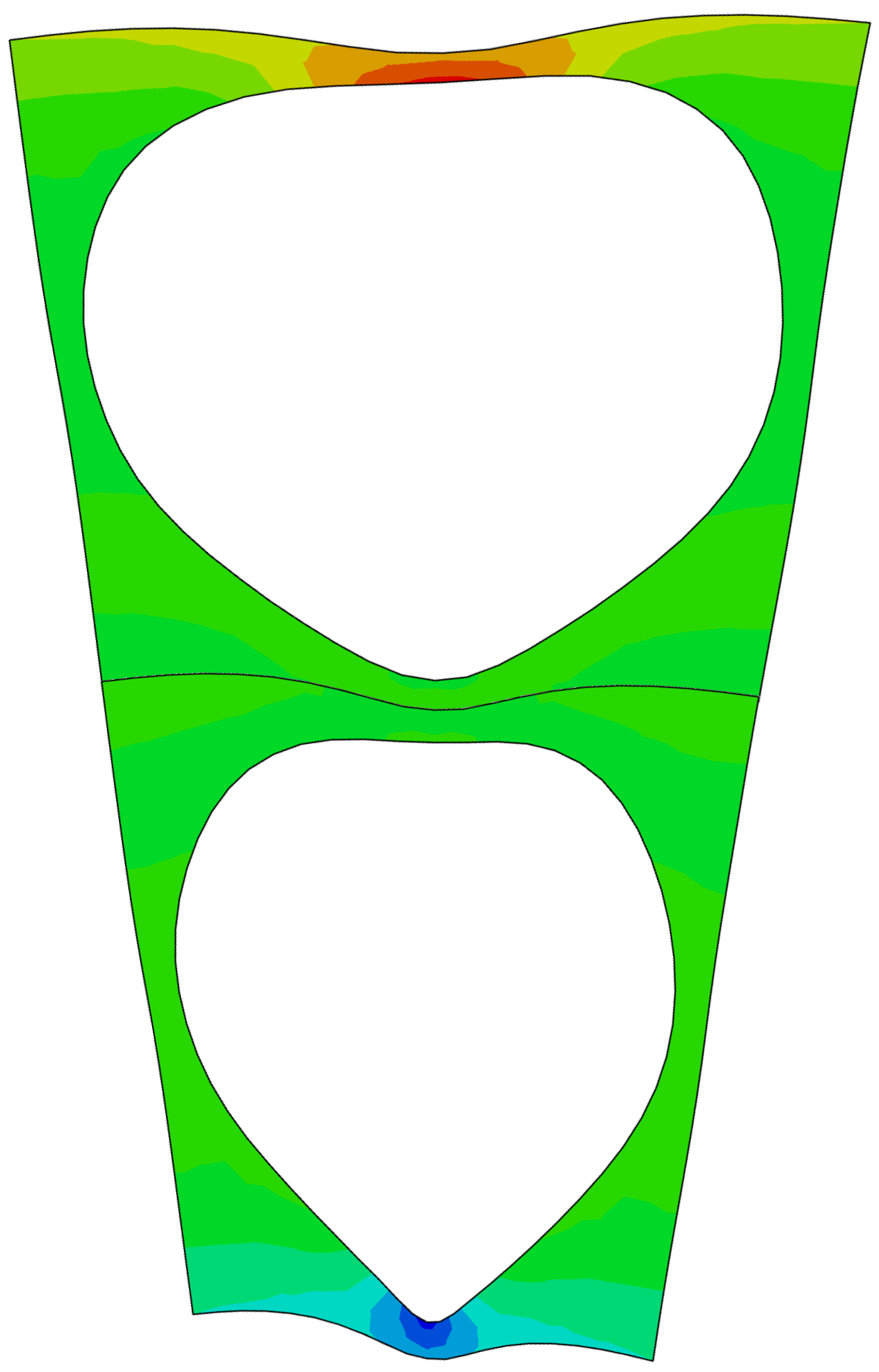}
        \caption{Micromorphic}
        \label{fig:ComparisonDeformation_Micromorphic}
        \end{subfigure}
        \caption{Comparison of the  microscale stresses computed by the different approaches in two-unit cell beam slices with $c\!=\!63.6\%$.}
        \label{fig:ComparisonDeformation}
\end{figure}

Firstly, the local fields at the microscale are examined in Fig.~\ref{fig:ComparisonDeformation}.
It shows the results from the direct numerical simulations in Fig.~\ref{fig:ComparisonDeformation_DNS} as a reference in comparison to respective FE\textsuperscript{2} simulations in Figs.~\ref{fig:ComparisonDeformation_Cauchy} to \ref{fig:ComparisonDeformation_Micromorphic}. Latter Figures have been generated by stacking the microscopic fields from macroscopic Gauss points at respective positions.
Classical homogenization in Fig.~\ref{fig:ComparisonDeformation_Cauchy} incorporates interactions of neighboring cells only via average strains, leading to signifcant deviations to the reference DNS solution in the considered example with low scale separation. 
In contrast, Cosserat and micromorphic theories incorporate additional bending-type deformation modes of the RVE leading to a significant improvement as can be seen in Figs.~\ref{fig:ComparisonDeformation_Cosserat} and \ref{fig:ComparisonDeformation_Micromorphic}. Even the warpage of the top and bottom ligaments is captured, though quantitatively slightly underestimated.
For this loading case of pure bending, the difference between Cosserat and micromorphic continua in Figs.~\ref{fig:ComparisonDeformation_Cosserat} and \ref{fig:ComparisonDeformation_Micromorphic} is only minor.

\begin{figure}
    \centering
    \begin{subfigure}{0.49\textwidth}
    \centering
    \nocite{McGregor2014}
    \includegraphics{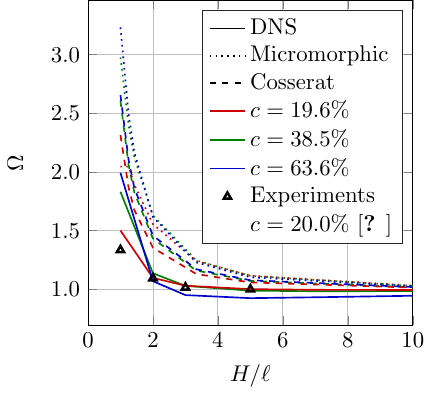}
    \caption{Linear elastic behavior}
    \label{fig:Comparison-elast}
    \end{subfigure}
    \begin{subfigure}{0.49\textwidth}
    \centering
    \includegraphics{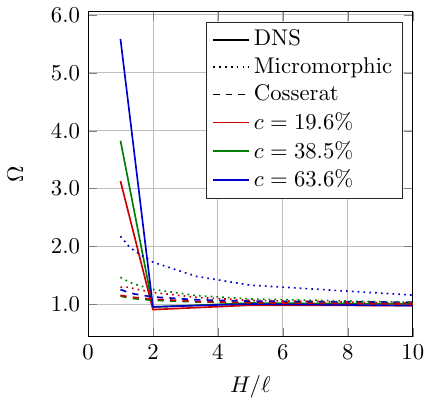}
    \caption{Limit moment for ideal plastic behavior}
    \label{fig:Comparison-plast}
    \end{subfigure}
    \caption{Size effects of the bending beam benchmark problem determined from DNS and direct FE\textsuperscript{2} simulations with different material properties.}
\end{figure}

Fig.~\ref{fig:Comparison-elast} shows the influence of the scale separation ratio $H/\ell$, i.~e., the number of unit cells over beam height, on the macroscopic bending stiffness for different values of relative porosity $c$. The results are presented in a normalized way as proposed by \citet{Gauthier1975}, and employed by Lakes and co-workers in numerous studies like \citep{Lakes1983,Lakes1986}, using the ratio
\begin{align}
\Omega&=\dfrac{M}{M^\text{Cauchy}}\ .\label{eq:Omegabend}
\end{align}
between the obtained bending moment $M$ to the respective expectation $M^\text{Cauchy}$ from classical Cauchy theory. For the elastic case, the reference solution amounts to $M^\text{Cauchy}\!=\!E^*I\,\gamma$ with the effective elastic Young's modulus $E^*$ of the material and second moment $I\!=\!B H^3/12$ of cross sectional area.

Firstly, it can be observed in Fig.~\ref{fig:Comparison-elast}, that there is only a minor difference between the predictions from micromorphic and Cosserat theories, as could have been anticipated from respective microscale results in Fig.~\ref{fig:ComparisonDeformation}. Compared to the fully resolved DNS, both continuum theories yield the same trend of a positive size effect (\enquote{smaller is stiffer}), but slightly overestimate this effect. Fig.~\ref{fig:Comparison-elast} contains also respective experimental data of \citet{McGregor2014} on a meta-material with exactly the ideal structure taken for the present DNS. That is why, these experimental data match quite well with the present DNS results at nearly some porosity of 20\%.

Remarkably, the present DNS even predicts a slight negative size effect for the highest investigated porosity of 63.6\% in the regime $4\!\leq\! H/\ell \!\leq\! 10$ as can be seen in Fig.~\ref{fig:Comparison-elast}.  Until now, it was thought that such negative size effects for foams are only due to the \enquote{edge effect} of incomplete cells at the specimen surface, cf.\ e.g.~\citep{Rueger2016,Wheel2015,Kirchhof2023} and references therein. However, this reason is excluded in the present study with totally closed cells. 
Rather, the high bending compliance of the closed struts at the specimen surface seems to be sufficient, cf.~Fig.~\ref{fig:ComparisonDeformation}.

\begin{figure}
    \begin{subfigure}{0.32\textwidth}
        \centering
        \includegraphics[width=0.95\textwidth]{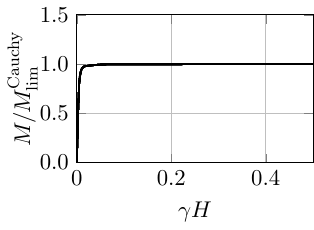}
        %\vspace*{-1.5ex}
        \caption{Cauchy}
        \label{fig:LimitLoadCauchy}
    \end{subfigure}\hfill
    \begin{subfigure}{0.32\textwidth}
        \centering
        \includegraphics[width=0.95\textwidth]{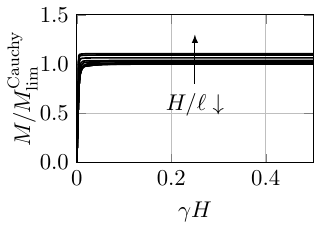}
        %\vspace*{-1.5ex}
        \caption{Cosserat}
        \label{fig:LimitLoadCosserat}
    \end{subfigure}\hfill
    \begin{subfigure}{0.32\textwidth}
        \centering
        \includegraphics[width=0.95\textwidth]{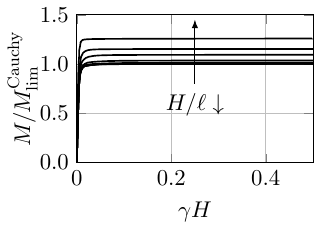}
        %\vspace*{-1.5ex}
        \caption{Micromorphic}
        \label{fig:LimitLoadFullMicromorph}
    \end{subfigure}
    \caption{Beam limit moment of the different theories for a porosity $c\!=\!38.5\%$ and ratios $H/\ell\!=\![2.0,3.33,5.0,10.0,20.0,40.0,50.0,66.66,100.0]$.}
    \label{fig:LimitLoad}
\end{figure}

For addressing inelastic behavior, ideal plasticity is incorporated in the material model of the bulk material. In this regime, it is convenient to assess the size effects with the ultimate load $M_{\text{lim}}$ of the beam, which is finally reached when increasing curvature $\gamma$ as displayed in Fig.~\ref{fig:LimitLoad} for the different theories on the example of porosity $c\!=\!38.5\%$ for different number $H/\ell$ of unit cells over beam height. As expected, ratio $H/\ell$ has no influence on load-deflection curves if Cauchy homogenization theory is applied in Fig.~\ref{fig:LimitLoadCauchy}. 
Rather, the limit moment amounts the the well-known value $M^{\mathrm{Cauchy}}_{\mathrm{lim}}\!=\!\sigma_\mathrm{F}^*\, BH^2/4$ in this case. Therein, $\sigma_\text{F}^*$ refers to the homogenized yield strength of the material identified from uniaxial tension simulations on the unit cell. 
In contrast, micromorphic theory as well as Cosserat theory both predict a positive size effect also in the plastic regime as can be seen in Figs.~\ref{fig:LimitLoadCosserat} and \ref{fig:LimitLoadFullMicromorph}, now even with a non-negligible difference between both theories.
Again the size effect is quantified in Fig.~\ref{fig:Comparison-plast} using dimensionless ratio $\Omega\!=\!M_{\mathrm{lim}}/ M^{\mathrm{Cauchy}}_{\mathrm{lim}}$ between ultimate load. Compared to the elastic case in Fig.~\ref{fig:Comparison-elast}, it is observed that the plastic limit load exhibits a lower size effect. With increasing porosity, the effect becomes noticeable, despite being of relevance with only minimal pores over the beam height. Furthermore, Cosserat theory as well as micromorphic theory predict smoother transitions in the regime of low scale separation $H/\ell$, as it is to be expected from continuum theories.

%---------------------------------------------------------------------------------------------------------------------------------------
\subsection{Plate with a hole}
\begin{figure}
    \centering
    \begin{subfigure}{0.3\textwidth}
      \centering
      \includegraphics[width=0.99\textwidth]{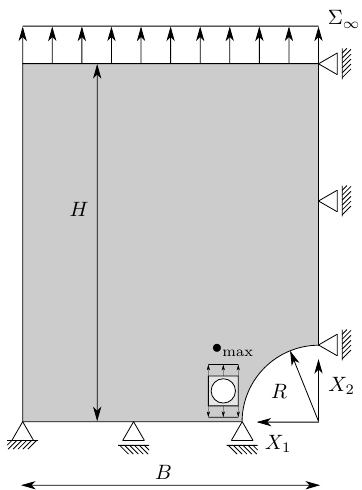}
      \caption{Scheme}
      \label{fig:SML-scheme}
    \end{subfigure}\hfill
    \begin{subfigure}{0.68\textwidth}
      \centering
      \includegraphics{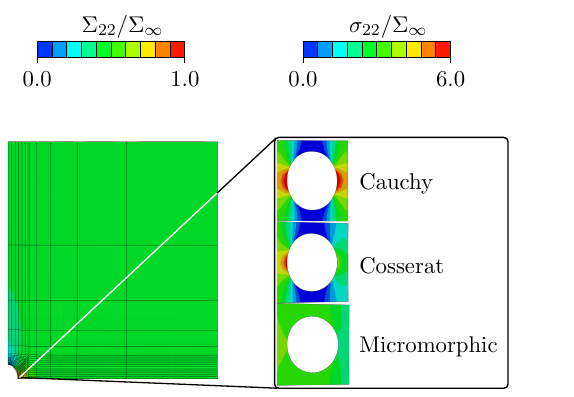}
      % %\def\svgwidth{0.7\textwidth}% without frame
      % \def\svgwidth{0.78\textwidth}% with frame
      % \begin{tikzpicture}
      %     \node[anchor=south west] at (0,0) {\import{figures/PlateWithHole/}{SML-results-frame.pdf_tex}};
      %     \node[anchor=south west] at (0.0,5.0) {\abqscaletenminmaxhorizontal{1.0}{0.0}{\hspace*{1em}$\Sigma_{22}/\Sigma_\infty$}{0.5}{1}};
      %      \node[anchor=south west] at (4.5,5.0) {\abqscaletenminmaxhorizontal{6.0}{0.0}{\hspace*{1em}$\sigma_{22}/\Sigma_\infty$}{0.5}{1}};
      %     \node at (9.5,5) {\phantom{a}};%for keeping text in figure
      %  \end{tikzpicture}
       \caption{Macroscopic and microscopic stress fields}
       \label{fig:SML-results}
    \end{subfigure}
\caption{Plate with a hole problem: Scheme and macroscopic and microscopic stress fields for different theories ($R/\ell\!=\!1$, $c\!=\!38.5\%$).}
\label{fig:SML}
\end{figure}

%---------------------------------------------------------------------------------------------------------------------------------------
A common numerical example is the plate with a hole. Its characteristic is the occurrence of a stress concentration %$\Sigma_{\text{max}}^{\text{Cauchy}}\!=\!3\,\Sigma_\infty$ 
$\Sigma_{\text{max}}^{\text{Cauchy}}/\Sigma_\infty$ 
at the hole for homogeneous far-field tension $\Sigma_\infty$, independent of the radius $R$ of the hole occurring in the loading direction on the side of the hole.
In the numerical implementation for the following investigations, a finite quarter model $H\!=\!B\!=\!20R$ with respective symmetry conditions is employed as sketched in Fig.~\ref{fig:SML-scheme}.  
At the nodes at top boundary, a displacement $U_\infty$ is applied in $X_2$ direction.
For simplicity, the value of the stress $\Sigma_{22}$ at the top left corner is taken as reference far-field stress $\Sigma_\infty$ for the following investigations
(which actually deviates slightly from the theoretical value $\Sigma_\infty\!=\!E^*U_\infty/H$ due to the finite dimensions of the FE model).
For  micromorphic and micropolar continua, the conditions $\chi_{12}\!=\!\chi_{21}\!=\!0$ and $\Phi_3\!=\!0$, respectively, are required at the lines of symmetry in addition to the conventional displacement condition, together with trivial natural boundary conditions for the remaining DOFs.
Typical results of respective FE\textsuperscript{2} simulation are shown in Fig.~\ref{fig:SML-results} for a low scale separation between radii of hole and pore\footnote{DNS are not used as reference here since the cut of a smooth hole into a periodic cubic structure could hardly be interpreted quantitatively in terms of macroscopic stresses, in particular for low scale separation.}.
In addition, the microscopic stress fields from the Gauss point closest to the hole are embedded at the right-hand side for the different theories.
Cauchy theory yields a symmetric distribution of microscopic stresses with further increased magnitude.
In contrast, Cosserat theory predicts a non-symmetric distribution of microscopic stresses due to the appearance of additional moment stresses in this region of high macroscopic gradients. 
Micromorphic theory on the other hand predicts a significantly lower stress level on the micro scale, which shows that the hole and the pores are hardly distinguishable as a result of the low scale separation. This behavior is expected to because the overall structure becomes more equally distributed in context of inhomogeneities, overruling the uniqueness of the hole and consequently its stress singularity on the hole vanishes.

\begin{figure}
    \centering
    \begin{subfigure}{0.49\textwidth}
        \centering
        \includegraphics{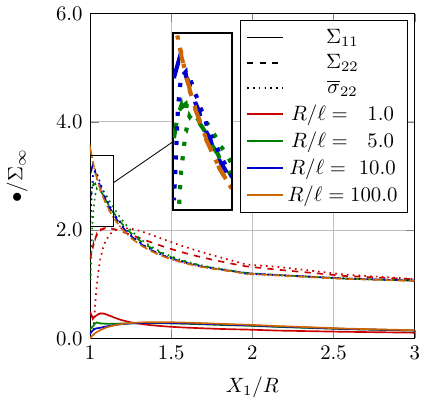}
        \caption{Ligament stresses $\Sigma_{11}$, $\Sigma_{22}$ and $\overline{\sigma}_{22}$}
        \label{fig:stresses_ligament_elastic}
    \end{subfigure}
    \hfill
    \begin{subfigure}{0.49\textwidth}
    \centering
        \includegraphics{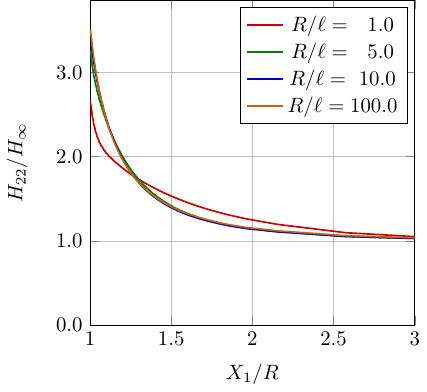}
        \caption{Ligament strain $E_{22}\!=\!H_{22}$}
        \label{fig:strains_ligament_elastic}
    \end{subfigure}
    \caption{Stresses and strains in the ligaments for a plate with a hole ($c\!=\!38.5\%$).}
    \label{fig:stressstrains_ligament_elastic}
\end{figure}

Fig.~\ref{fig:stressstrains_ligament_elastic} takes a closer look at the distributions of stresses and strains in the ligament for different values of scale separation ratio $R/\ell$. 
It is found that the concentration of hoop stresses and strains decreases with decreasing values of $R/\ell$,
i.e., the more similar the pore and hole sizes become. In this case, also the difference $s_{22}\!=\!\Sigma_{22}-\overline{\sigma}_{22}$ increases due to the presence of hyperstresses according to higher-order balance~\eqref{eq:angularmomentum}.
Vice versa, the classical Cauchy solution is obtained asymptotically for $R/\ell\rightarrow\infty.$
This behavior is not only plausible from a physical point of view, but it is also in agreement with previous results for the Cosserat continuum \citep{Neuber1966,Neuber1966a} and for a microstrain continuum \citep{Dillard2006}.

\begin{figure}
    \centering
    \begin{subfigure}{0.49\textwidth}
        \centering
        \includegraphics{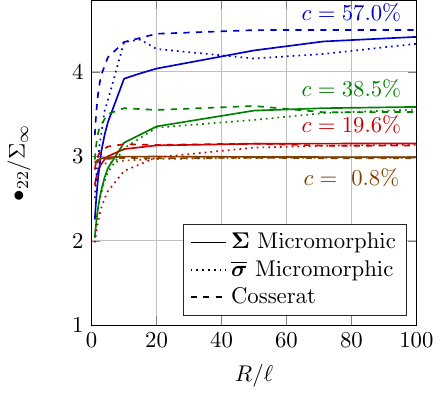}
        \caption{}
        \label{fig:PWH-elast}
    \end{subfigure}
    \begin{subfigure}{0.49\textwidth}
      \includegraphics{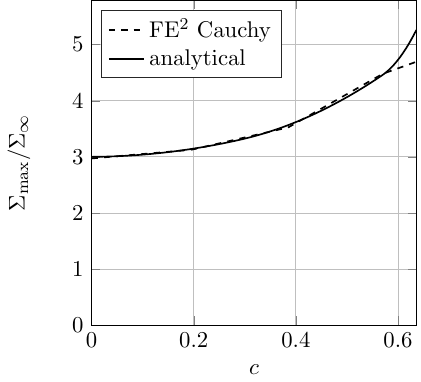}
      \caption{}
      \label{fig:Kt-c}
    \end{subfigure}
    \caption{Maximum macroscopic hoop stress for a plate with a hole in linear elastic regime: \subref{fig:PWH-elast} numerical FE\textsuperscript{2} results for micromorphic theory and \subref{fig:Kt-c} comparison of asymptotic value to analytical solution.}
\end{figure}

In contrast to these studies with phenomenological constitutive relations, the present homogenization approach allows to investigate the effect of the microstructure explicitly. In this sense, the maximum stress at the ligament is evaluated in Fig.~\ref{fig:PWH-elast} for different values of relative porosity $c$ (note it does not appear directly at the surface of the hole for smaller scale separation ratios).
The results show that the size effect increases with increasing porosity $c$, both, regarding amplitude as well as respective range of scale separation ratio $R/\ell$. But the size effect vanishes for microhomogeneous material $c\rightarrow0$. The latter prediction is plausible, but it is has not been attained by a number of homogenization approaches towards generalized continua. 
Regarding the asymptotic values of stress concentration, one might be wondering at first glimpse why $\Sigma_{\mathrm{max}}/\Sigma_{\infty}(R/\ell\rightarrow\infty)\approx3.0$ is reached only for the very low porosity $c\!=\!0.8\%$. In this context, it is recalled that the employed unit cell has only cubic symmetry and the Zener anisotropy ratio increases with $c$ as exploited in Fig.~\ref{fig:Zenerratio}.
The analytical solution for the stress concentration in an anisotropic plate with a circular hole \citep{Lekhnitskii1968anisotropic} can be written for cubic symmetry  in terms of the Zener ratio $\alpha_r$ as
\begin{align}
    \dfrac{\Sigma_{\mathrm{max}}}{\Sigma_{\infty}}=1+\beta_1+\beta_2\qquad\text{with}\quad \beta_{1,2}=\sqrt{p\pm\sqrt{p^2-1}}\ ,\quad p=\dfrac{1+(1-\alpha_r)\overline{\nu}_{\mathrm{eff}}}{\alpha_r}\ .
    \label{eq:notch_intensity_analytical}
\end{align}
Therein, the effective Poisson's ratio is computed from the aforementioned Lamé-type constants via $\overline{\nu}_{\mathrm{eff}}\!=\!\bar{K}_{\mathrm{eff}}/(\bar{K}_{\mathrm{eff}}+2\bar{\mu}_{\mathrm{eff}}^{\mathrm{longitud.}})$ with $\bar{K}_{\mathrm{eff}}\!=\!\bar{K}_{E}-\bar{K}_{E\chi}^2/\bar{K}_{\chi}$. 
Fig.~\ref{fig:Kt-c} compares the numerically obtained asymptotic values of ${\Sigma_{\mathrm{max}}}/{\Sigma_{\infty}}$, corresponding to Cauchy theory, with the analytical solution~\eqref{eq:notch_intensity_analytical}. A good agreement is found. Only for the highest porosity $c\!=\!63.6\%$ a slight deviation occurs as it becomes difficult to construct a robust mesh in this case where the ligament at the microscale between pore surface and unit cell edge is very narrow. 

\subsection{Flow-through filter}
\begin{figure}[hbt]
    \centering
    \includegraphics[width=0.75\textwidth]{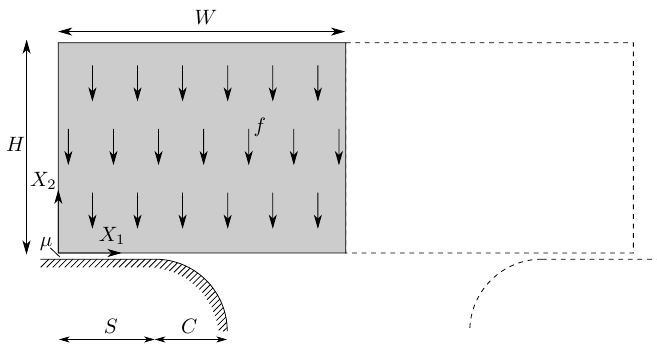}
    \caption{Schematic of the flow-through filter model with dimensions $W/H\!=\!3/2$, $S\!=\!1/3\,W$, $C\!=\!1/4\,H$, and environmental conditions comprising friction coefficient $\mu\!=\!0.3$ and body force $f\!=\!0.02\, E/H$, where $E$ is the elastic modulus of the bulk material.}
    \label{fig:pressureloadedfilter}
\end{figure}

Finally, a viscoplastic flow-through filter with a support on the outer domain, respecting friction, is investigated as illustrated in Fig.~\ref{fig:pressureloadedfilter}. 
Symmetry is exploited to reduce computational expense. This boundary value problem is inspired by the application of ceramic foam filters in the cleaning of steel melts during production, which typically have a low scale separation between height and pore size. The aim is to be able to predict size effects in tailoring the structural response of these filters. The loading of the filter results from its resistance to the melt flow, which is modeled here as a volumetric body force $f$, cf.~\citep{lange+etal22}. In the corresponding DNS, a statically equivalent volume force $f_{\mathrm{bulk}}\!=\!f/(1-c)$ is applied to the bulk material only. Self-contact of pore surfaces is taken into account in DNS as well as in direct FE\textsuperscript{2} simulations.

At the high temperatures experienced during the production processes, the ceramic bulk material becomes ductile and begins to creep.
This behavior is captured here by means of a strain hardening power law
\begin{align}
    \tens{d}^\mathrm{cr}
    &=\dfrac{3}{2}\,\dfrac{\dev{\tens{\sigma}}}{\sigma_\mathrm{eq}}\,\dot{\varepsilon}^\mathrm{cr}_\mathrm{eq}\,\quad \text{with}\ \
    \dot{\varepsilon}^\mathrm{cr}_\mathrm{eq}=\left(\dfrac{\sigma_\mathrm{eq}}{A}\right)^{\frac{n}{m+1}}\left[(m+1)\,\varepsilon^\mathrm{cr}_\mathrm{eq}\right]^{\frac{m}{m+1}}%\ ,\ \\sigma_\mathrm{eq}&=\sqrt{3J_2}\ ,
    \label{eq:creep_law}
\end{align}
for the creep part of the Eulerian rate of deformation tensor $\tens{d}^\mathrm{cr}$,
in terms of the equivalent von Mises (Cauchy) stress. Therein, the independent parameters are stress factor $A$, stress exponent $n$, and exponent $m$. Their respective values $A\!=\!22.09\,E\,t_{\mathrm{ref}}^{m+1}$, $n\!=\!1.06$, $m\!=\!-0.56$, with $t_{\mathrm{ref}}\!=\!1\,\mathrm{s}$, and Young's modulus $E$ are taken from \citep{lange+etal22}. Further details on the material law~\eqref{eq:creep_law} and its implementation can be found within this reference, too.
Furthermore, a representative filter porosity of $c\!=\!38.5\%$ is assumed for the following investigations.

Since finite deformations are considered, the present implementation cannot be applied to the isolated Cosserat theory. The required polar decomposition of $\tens{\Phi}$ would result in nonlinear equations, that are not straightforwardly implementable in the direct FE\textsuperscript{2} framework. By the fact that the micromorphic theory includes Cosserat theory as a special case, the effort is avoided here in favor of the complete micromorphic theory.

\begin{figure}
    \centering
    % \def\svgwidth{0.99\textwidth}
    % \begin{tikzpicture}
    % \node[anchor=south west] at (0,0) {
    % \import{figures/PressureLoadedFilter/}{PLF_nlgeom_results-frame.pdf_tex}};
    % % without frame
    % % \node[anchor=south west] at (0.25,7.1) {\abqscaletenminmaxhorizontal{-0.23}{0.03}{$\chi_{22}$}{0.5}{1}};
    % % \node[anchor=south west] at (7.0,7.1) {\abqscaletenminmaxhorizontal{1.2}{-3.8}{$\sigma_{22}H/(E_{\mathrm{eff}}|U_2|)$}{0.5}{1}};
    % \node[anchor=south west] at (0.25,6.8) {\abqscaletenminmaxhorizontal{-0.23}{0.03}{$\chi_{22}$}{0.5}{1}};
    % \node[anchor=south west] at (7.5,6.8) {\abqscaletenminmaxhorizontal{1.2}{-3.8}{$\sigma_{22}H/(E_{\mathrm{eff}}|U_2|)$}{0.5}{1}};
    % \draw[black] (11.75,14.8)--(12.25,14.8) node[right] {DNS};
    % \draw[black,dashed] (11.75,14.3)--(12.25,14.3) node[right] {Cauchy};
    % \draw[black,dotted,thick] (11.75,13.8)--(12.25,13.8) node[right] {Micromorphic};
    % \end{tikzpicture}
    \includegraphics{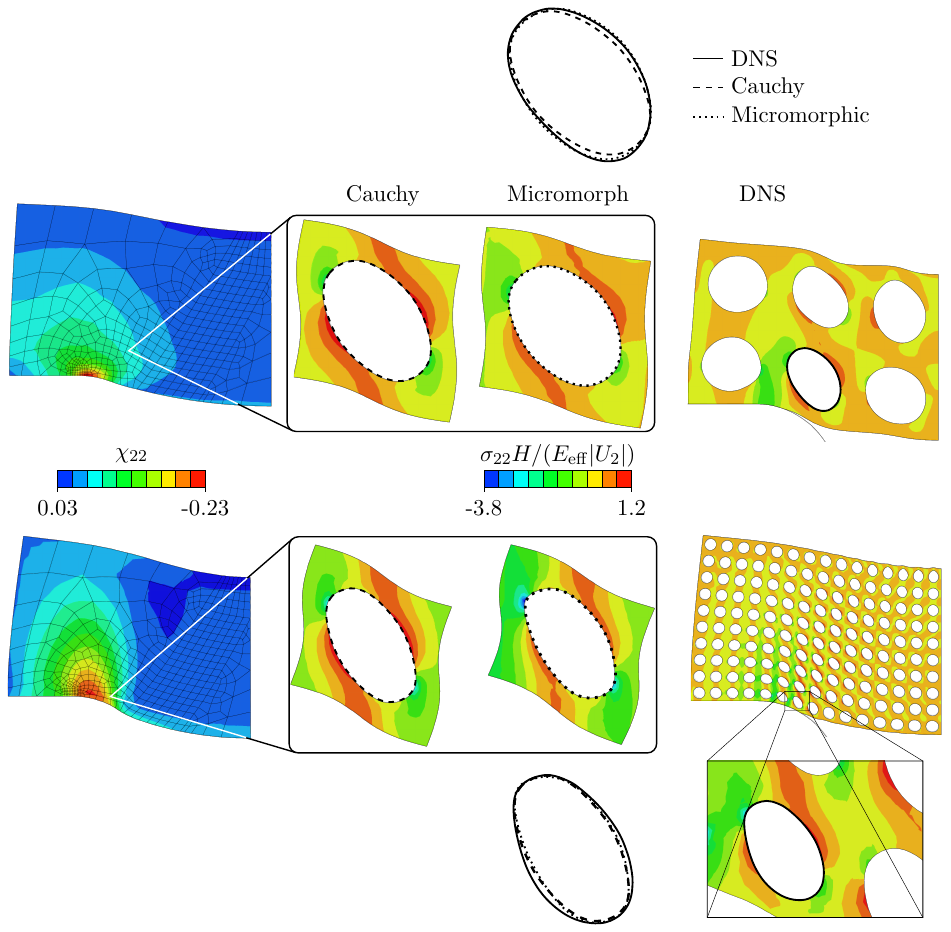}
    \caption{Comparison of the different modeling approaches for the flow-through filter behavior with $H/\ell\!=\!2$ (top) and $H/\ell\!=\!10$ (bottom) unit cells over the height, with results highlighted at the location of the maximally distorted pore. Displayed are the macroscopic microdeformation component $\chi_{22}$ and microscopic stress $\sigma_{22}$, as well as the superimposed pore contours for an applied load of $f\!=\!0.014\, E/H$.}
    \label{fig:PLF_Comparison}
\end{figure}

Firstly, Fig.~\ref{fig:PLF_Comparison} shows the predicted macroscopic and microscopic deformations and stresses in the initial (essentially) linear elastic loading regime, for two different relative pores size microstructures.
In particular, special focus is given to the unit cells located in the highly stressed region in the vicinity of the support contact area.
For comparison with the FE\textsuperscript{2} simulations, the cell locations have been chosen to correspond to the macroscopic Gauss points at the respective centers of the maximally distorted pores, as identified in the DNS. 
At first glimpse, the microscopic deformations and stress fields look similar for the Cauchy and micromorphic FE\textsuperscript{2} simulations, even in the case of only $H/\ell\!=\!2$ pores over the height of the structure, as visualized in the upper row of Fig.~\ref{fig:PLF_Comparison}.
In detail, however, the micromorphic simulation is observed to predict lower stresses directly at the pore surface and an asymmetric distribution of stresses between opposite corner regions of the unit cell, in contrast to Cauchy theory but in agreement with a respective DNS. However, the local stress magnitudes are quantitatively overestimated by the micromorphic theory as well.
The improved prediction by the micromorphic theory can also be found when comparing the contours of the deformed pore in the very top row of Fig.~\ref{fig:PLF_Comparison}.
For the higher scale separation of $H/\ell\!=\!10$ pores over height, hardly any difference can be found between Cauchy theory and micromorphic theory, neither for microscopic stress fields nor for the deformation of the highly stressed unit cell.
Note that the latter have different locations of center for different values $H/\ell$ which is why the deformed unit cells in column \enquote{Cauchy} are different, although both respective FE\textsuperscript{2} simulations yield identical results.

\begin{figure}[hbt]
    \centering
        \includegraphics{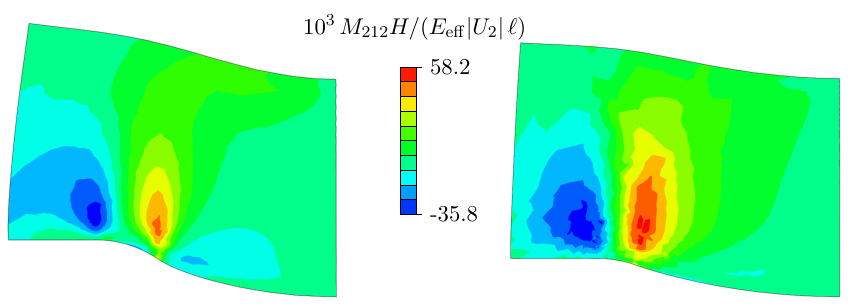}
        % \begin{tikzpicture}
        %     \node[anchor=south west] at (-6.5,0) {\includegraphics[width=0.35\textwidth]{figures/PressureLoadedFilter/FM-RVE2_M212_130_-80.png}};
        %     \node[anchor=south west] at (2,0) {\includegraphics[width=0.35\textwidth]{figures/PressureLoadedFilter/FM-RVE10_M212_650_-400.png}};
        %     \node[anchor=south west] at (-1.75,1.15) {\abqscaletenminmax{58.2}{-35.8}{$10^3\,M_{212}H/(E_{\mathrm{eff}}|U_2|\,\ell)$}{0.5}{1}};
        % \end{tikzpicture}
    \caption{Hyperstress component $M_{212}$ for $H/\ell\!=\!10$ (left) and $H/\ell\!=\!2$ (right) unit cells over the filter height.}
    \label{fig:PLF-hyperstress}
\end{figure}

Fig.~\ref{fig:PLF-hyperstress} shows the distribution of hyperstresses from the micromorphic simulation for both cases. 
In the case of low scale separation $H/\ell\!=\!2$, the magnitudes reach significant values over the complete height. But even in the case of higher scale separation $H/\ell\!=\!10$, strong hyperstress fields are \enquote{emitted} from the support contact region.
In this context it is recalled, that the scale separation between radius $R$ of support and intrinsic length still amounts to $R/\ell\!=\!2.5$ only (even for such a practically unrealistic large ratio $R/H$). 
 
\begin{figure}
    \centering
    \includegraphics{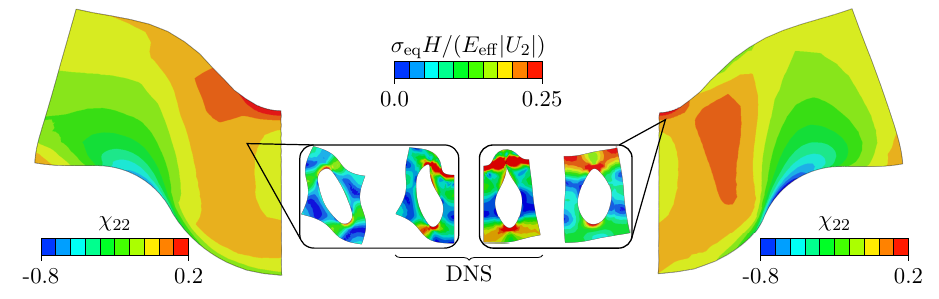}
    % \begin{tikzpicture}
    %     \node[anchor=south west] at (-8.0,0) {\includegraphics[width=0.45\textwidth]{figures/PressureLoadedFilter/CreepLocalizationRVE10-frame.pdf}};
    %     \node[anchor=south east] at (7.0,0) {\includegraphics[width=0.45\textwidth]{figures/PressureLoadedFilter/CreepLocalizationRVE2-frame.pdf}};
    %     \node[anchor=south west] at (-2.4,2.75) {\abqscaletenminmaxhorizontal{0.25}{0.0}{$\sigma_{\mathrm{eq}}H/(E_{\mathrm{eff}}|U_2|)$}{0.5}{1}};
    %     \node[anchor=south west] at (-8.45,-0.25) {\abqscaletenminmaxhorizontal{0.2}{-0.8}{$\chi_{22}$}{0.5}{1}};
    %     \node[anchor=south east] at (7.45,-0.25) {\abqscaletenminmaxhorizontal{0.2}{-0.8}{$\chi_{22}$}{0.5}{1}};
    %     %\node[] at (-1.125,-0.75) {DNS};
    %     \draw[decoration={brace,mirror,raise=0.5cm},decorate] (-1.75,1.0)--(0.75,1.0) node[pos=0.5,anchor=north,yshift=-0.55cm] {DNS};
    % \end{tikzpicture}
    \caption{Microdeformation component $\chi_{22}$ at $U_2/H\!=\!-0.65$ with preceded creeping for $H/\ell\!=\!2$ (left) and $H/\ell\!=\!10$ (right) unit cells over the height, with corresponding RVE of the upper filter pore for comparison to the DNS and at the location of maximum  $\chi_{22}$.}
    \label{fig:PLF-creep-chi22}
\end{figure}

When the body force $f$ is increased and finally kept constant during the hold time, the deflections increase continuously, including all classic and non-classic deformation measures. Fig.~\ref{fig:PLF-creep-chi22} shows the microdeformation fields for two scale separation ratios $H/\ell$. 
It becomes clear, that the microdeformation increased as expected, if one compares it to the elastic state displayed in Fig.~\ref{fig:PLF_Comparison} without creeping. Furthermore, selected unit cells of the micro scale are shown for a comparison with the DNS (lower row in Fig.~\ref{fig:PLF-creep-chi22} with Mises stress $\sigma_{\mathrm{eq}}$ as local fields. Again, the excerpt unit cells from FE\textsuperscript{2} corresponds to the integration point of the macroscale where the center of the pore lies in the DNS. Due to symmetry boundary conditions, the DNS unit cells are showing a straight edge in the middle of the filter. The corresponding RVEs from the micromorphic investigation are further away from the symmetry boundary and thus not affected by these exactly. However, the micromorphic theory is capturing the deformation modes of the unit cell in qualitatively good agreement, but with certain quantitative deviations. 

\begin{figure}
    \centering
    \begin{subfigure}{0.49\textwidth}
        \centering
        \includegraphics{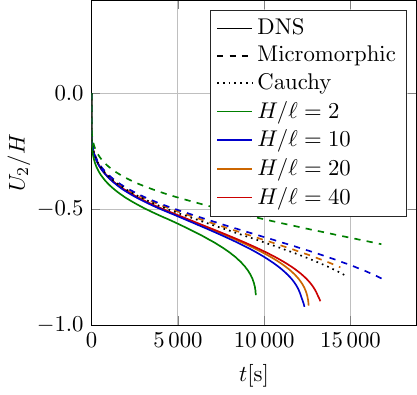}
        \caption{Maximum deflection of filter bottom}
        \label{fig:TertiaryCreepDeflection}
    \end{subfigure}
    \begin{subfigure}{0.49\textwidth}
        \centering
        \includegraphics{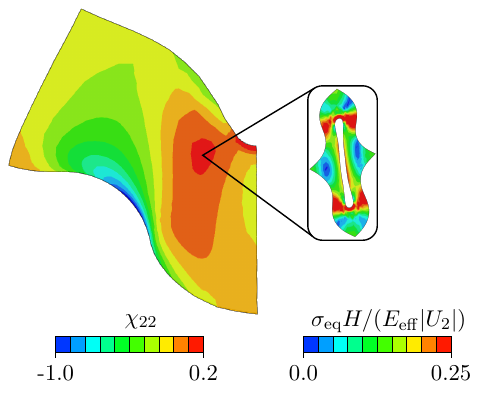}
        % \begin{tikzpicture}
        %     \node[anchor=south west] at (0,1.25) {\includegraphics[width=0.8\textwidth]{figures/PressureLoadedFilter/CreepLocalizationRVE2_tmax-frame.pdf}};
        %     \node[anchor=south west] at (0.25,0) {\abqscaletenminmaxhorizontal{0.2}{-1.0}{\hspace*{1em}$\chi_{22}$}{0.5}{1}};
        %     \node[anchor=south west] at (4.5,0) {\abqscaletenminmaxhorizontal{0.25}{0.0}{\hspace*{1em}$\sigma_{\text{eq}}H/(E_{\mathrm{eff}}|U_2|)$}{0.5}{1}};
        % \end{tikzpicture}
        \caption{Localization for tertiary creep}
        \label{fig:TertiaryCreepRVE}
    \end{subfigure}
    \caption{Creep curve with tertiary creep and difference strain component $\chi_{22}$ for $10$ unit cells of porosity $c\!=\!38.5\%$ over the flow-through filter height after $7\,\mathrm{h}$ creeping with localization and corresponding microscale deformation.}
    \label{fig:TertiaryCreep}
\end{figure}

Fig.~\ref{fig:TertiaryCreepDeflection} shows the resulting macroscopic creep curves predicted by the three simulation approaches for different scale separation ratios $H/\ell$, i.e., numbers of pores over height. Firstly, it can be observed that the three characteristic regions of primary, secondary and tertiary creep occur. The secondary regime stem from the creep law~\eqref{eq:creep_law} of the bulk material. In contrast, the primary and tertiary range is a result if the hierarchical structure of a foam. While the primary creep regime results from an internal redistribution of stresses, the predicted final tertiary creep arises from the large deformations that accumulate. Thus, leading to a reduced stress-carrying cross section at the microscale as can be seen in Fig.~\ref{fig:TertiaryCreepRVE}, where the pore is nearly collapsed. 

Furthermore, it can be observed in Fig.~\ref{fig:TertiaryCreepDeflection}, that DNS as well as micromorphic FE\textsuperscript{2} simulations converge to a common curve with increasing scale separation $H/\ell$.

Despite the principal description of local stress and deformation fields of the microstucture, the investigated structure shows a negative global size effect. For foams such a phenomenon was detected in multiple experimental and numerical \cite{Brezny1990,Liebold2016a,Frame2018} and simulative \cite{Ameen2018,Kirchhof2023} investigations. However, the micromorphic theory can describe positive size effects only. To capture negative size effects, an alternative continuum theory like stress gradient theory \cite{Forest2012,Huetter2020} is required. Unfortunately, this theory can only represent negative size effects. The second-order micromorphic continuum is able to capture positive and negative size effects, for which no homogenization theory is available to the authors knowledge.

\subsection{Indentation}
\begin{figure}[hbt]
    \centering
    \includegraphics[width=0.6\textwidth]{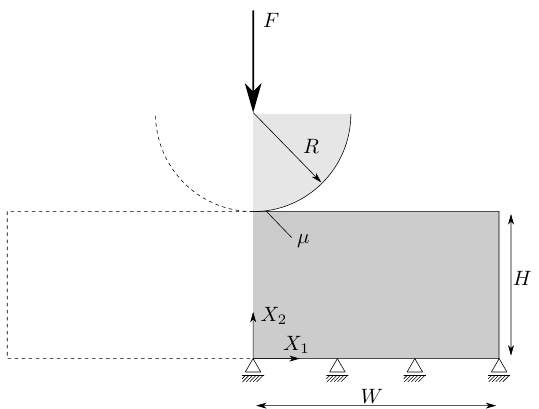}
    \caption{Scheme of the spherical indenter model with dimensions $W/H\!=\!5/3$, $R\!=\!1/3\,H$, loaded with punch force $F$ and friction coefficient $\mu$ between punch and specimen.}
    \label{fig:indenterscheme}
\end{figure}

Indentation is a common technique to quantify local plastic properties of a material. However, size effects can occur in this case as the scale separation between indenter and foam microstructure often is relatively small \citep{Andrews2001,Forest2005a,Tekoglu2011}.
To study this influence, the problem of a (plane) Brinnel indenter is considered, as schematically shown  in Fig.~\ref{fig:indenterscheme}. 
Again, symmetry is exploited to reduce computational costs.
Mises plasticity with linear hardening 
\begin{align}
    \sigma_{\mathrm{y}}&=\sigma_{\mathrm{y0}}+E^{\mathrm{H}}\varepsilon^{\text{pl}}_{\text{eq}}\ ,
\end{align}
in terms of the equivalent plastic strain $\varepsilon^{\text{pl}}_{\text{eq}}$ is utilized for the bulk material. In particular, values $\sigma_{\mathrm{y0}}\!=\!10^{-3}\,E$  and $E^{\text{H}}\!=\!3/2\cdot10^{-4}E$ are assumed for the initial yield stress and hardening modulus, respectively.
Finite deformation theory is employed, as considerable plastic deformations occur in the contact region. Isolated Cosserat theory therefore cannot be employed for the aforementioned reason, so that the focus is placed on comparing predictions of the complete micromorphic theory to DNS.

\begin{figure}
    \centering
   % \begin{subfigure}{0.49\textwidth}
    %    \centering
        %\input{figures/IndenterU2-force.tex}
        \includegraphics{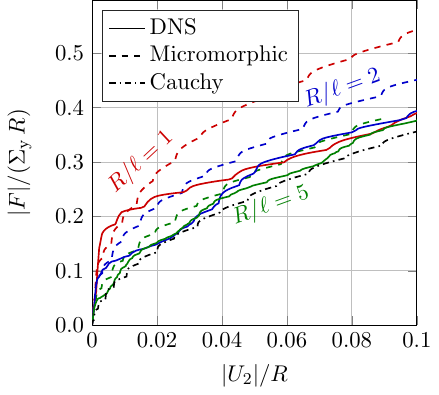}
        \caption{Vertical punch force vs.~displacement responses predicted for the indentation problem.}
        %\label{fig:punchforcedisplacement}
    %\end{subfigure}
    %\begin{subfigure}{0.49\textwidth}
     %   \centering
        %\input{figures/IndenterU2.tex}
        %\includegraphics{tikzfigures/IndenterU2.pdf}
        %\caption{Vertical contact stress and displacement components of the indentation model.}
        %\label{fig:punchforcedisplacement}
    %\end{subfigure}
    %\caption{Force indentation curves.}
    \label{fig:punchforcedisplacement}
\end{figure}

Corresponding indentation curves are shown in Fig.~\ref{fig:punchforcedisplacement} in a normalized way, for different levels of scale separation $R/\ell$. 

Note that $F$ therein represents the force per unit thickness. 
Firstly, it can be seen that the DNS with largest investigated scale separation ratio $R/\ell\!=\!5$ complies well with the respective Cauchy FE\textsuperscript{2} simulation, the latter also representing the limiting case $R/\ell\rightarrow\infty$ for the micromorphic theory.
Furthermore, the micromorphic theory predicts a positive size effect, i.~e., small specimens exhibit a higher normalized load-deflection curves in agreement with the DNS.
Quantitatively, the micromorphic theory predicts the initial part of the load-deflection curves quite accurately, but the subsection evolution of reaction forces is overestimated for low scale-separation ratios $R/\ell\!=\!2$ and $R/\ell\!=\!1$.

\begin{figure}
    \centering
    \includegraphics{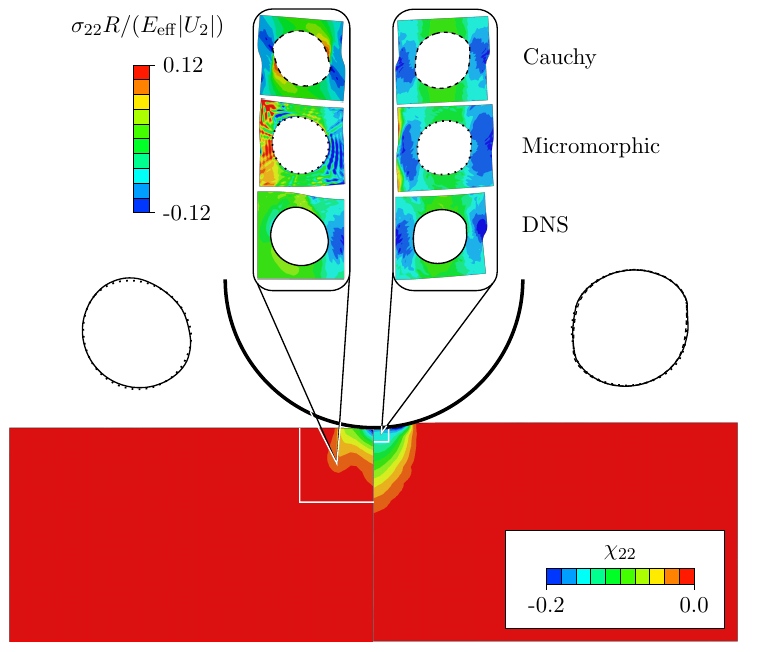}
    % \def\svgwidth{0.8\textwidth}
    % \begin{tikzpicture}
    %     \node[anchor=south east] at (0,0) {\import{figures/Indenter/}{Indenter-Comparison-frame.pdf_tex}};
    %     \node[anchor=south east] (chiscale) at (-0.8,0.4) {\abqscaletenminmaxhorizontal{0.0}{-0.2}{$\chi_{22}$}{0.5}{1}};
    %     \draw[black,fill=white] (chiscale.north west) rectangle (chiscale.south east);
    %     \node[anchor=south east] at (-0.8,0.4) {\abqscaletenminmaxhorizontal{0.0}{-0.2}{$\chi_{22}$}{0.5}{1}};
    %     \node[anchor=south east] at (-9.0,7.05) {\abqscaletenminmax{0.12}{-0.12}{$\sigma_{22}R/(E_{\mathrm{eff}}|U_2|)$}{0.5}{1}};
    % \end{tikzpicture}
    \caption{Comparison of local fields at the micro- and macroscales for the different modeling approaches and scale separations of $R/\ell\!=\!1$ (left) and $R/\ell\!=\!5$ (right), visualized at $|U_2|/R\!=\!0.03$.}
    \label{fig:indenter-comparison}
\end{figure}

The macroscopic fields of microdeformations as well as the microscopic fields in the highly stressed unit cell directly underneath the indenter are shown in Fig.~\ref{fig:indenter-comparison} for two different ratios of scale separation.
Regarding the field of microdeformation $\chi_{22}$ predicted by the micromorphic theory, a wider region of influence is found for the case of low scale separation at the left-hand side of Fig.~\ref{fig:indenter-comparison} compared to $R/\ell\!=\!5$ at the right-hand side, but with higher magnitudes of $\chi_{22}$ for low scale separation $R/\ell\!=\!1$. Furthermore, the region of highest values of $\chi_{22}$ is more localized for $R/\ell\!=\!1$ than for a larger scale separation $R/\ell\!=\!5$.
The reason for this behavior can be found at the microscale. The embedded unit cells on Fig.~\ref{fig:indenter-comparison} show that the micromorphic FE\textsuperscript{2} approach can predict the horizontal gradient in the stress field and thus the evolution of pore shape more accurately than classical Cauchy FE\textsuperscript{2} approach, compared to DNS though not in every detail of course.

\section{Summary and Conclusion}\label{sec:summary}
Foam, or foam-like, materials are often used in applications in which the separation between the component scale and the mean distance of pores is relatively low, so that size effects are observed. Generalized continuum theories have been used in numerous studies to model and predict these size effects at the macroscopic scale, albeit almost exclusively within the realms of phenomenological constitutive laws and elastic behavior.

In contrast, the present study has employed a micromorphic homogenization theory within an FE\textsuperscript{2} framework that includes Cosserat (micropolar) theory as a special case. In a first step, unit cell computations of the microstructure have been considered. The respective results do not only verify the correct implementation of the non-classical contributions, but additionally allow us to draw a connection to the constitutive parameters of established phenomenological laws and their interpretation at the microscale. 

Subsequently, the commonly used benchmark problems of beam bending and plate with a hole have been considered in the elastic and in the plastic regime, for both the Cosserat and the complete micromorphic theory.
In all cases, the results comply with previous positive size effects reported in studies on Cosserat and micromorphic approaches in literature. In contrast to those studies, the present homogenization approach yields \emph{predictions} on the influences of relative porosity $c$ and pore distance $\ell$ as measurable geometric microscale parameters. In this regard, our approach predicts correctly an increasing size effect with increasing $c$, which vanishes for the microhomogeneous material $c\rightarrow0$. Reasonable predictions have also been obtained for the influence of $\ell$, although in bending the size effect is slightly overestimated compared to DNS.

Furthermore, size effects in the irreversible behavior of foams under complex loading conditions and finite deformation kinematics have been simulated for the creep response of a flow-through filter and for plastic indentation.
In both cases, the micromorphic theory gives improved predictions of the local deformations at the microscale. For the indentation problem, also the positive size effect is predicted correctly for the force-displacement curves. In contrast, the DNS of the creep problem yields a negative size effect, which cannot be predicted by the micromorphic theory at all. More general continuum theories are necessary to address this effect.

In summary, the micromorphic FE\textsuperscript{2} framework has been found to be a universal and powerful approach to predict positive size effects in the reversible and irreversible deformation behavior of foam-like materials, while preserving all relevant microscale information.
Future studies need to address the question of whether this conclusion also holds  for more realistic 3D foam geometries. Moreover, the effect of the slightly different approaches from the literature for the micro-macro-coupling of micro deformation $\tens{\chi}$ and its gradient $\tens{K}$ on the predicted size effect, cf.~\citep{Biswas2017}, require further investigation.

\clearpage
% Acknowledgements
\medskip
\textbf{Acknowledgements} \par 
\noindent This work was funded by the Deutsche Forschungsgemeinschaft (DFG, German Research Foundation) -- Project-ID 169148856 -- CRC 920: Multi-Functional Filters for Metal Melt Filtration -- A Contribution towards Zero Defect Materials. Furthermore, the authors acknowledge computing time on the Compute Cluster of the Faculty of Mathematics and Computer Science at Technische Universit\"at Bergakademie Freiberg, operated by the Computing Center (URZ) and funded by the Deutsche Forschungsgemeinschaft (DFG, German Research Foundation) -- Project-ID 397252409.

% Conflict of interest
\medskip
\textbf{Conflicts of Interest} \par
\noindent The authors declare that there is no conflict of interest. 

%Appendix
\setcounter{section}{0}
\renewcommand{\thesection}{Appendix \Alph{section}}
\setcounter{equation}{0}
\renewcommand{\theequation}{\Alph{section}\arabic{equation}}

\section{Stress Measures in a Large Deformation Analysis}\label{sec:stressconversion}

Within this section, stress measures of first and second Piola-Kirchhoff type are marked by superscripts \enquote{PK1} and \enquote{PK2}, respectively. The respective stress measures can be converted as
\begin{align}
    \overline{\tens{\sigma}}^{\mathrm{PK2}}&=\tens{\Sigma}^{\mathrm{PK1}}\cdot\tens{F}^{-T}-(\tens{e}^{\mathrm{GL}}+\tens{I})\cdot\left(\tens{s}^{\mathrm{PK1}}-\tens{M}^{\mathrm{PK1}}:\tens{K}^{T(1\rightarrow3)}\right)^T\cdot\tens{F}^{-T}\\
    \tens{s}^{\mathrm{PK2}}&=\tens{\chi}^T\cdot\left(\tens{s}^{\mathrm{PK1}}-\tens{M}^{\mathrm{PK1}}:\tens{K}^{T(1\rightarrow3)}\right)\cdot(\tens{e}^{\mathrm{GL}}+\tens{I})^{-T}\\
    \tens{M}^{\mathrm{PK2}}&=\tens{\chi}^T\cdot \tens{M}^{\mathrm{PK1}}\ .
\end{align}
Therein, the product $\tens{M}^{\mathrm{PK1}}:\tens{K}^{T(1\rightarrow3)}$ is to be read as $M^{\mathrm{PK1}}_{iKL}K_{jKL}\!=\!M^{\mathrm{PK1}}_{iKL}\chi_{jK,L}$. As in conventional theory, $\overline{\tens{\sigma}}^{\mathrm{PK2}}$ is symmetric, whereas $\overline{\tens{\sigma}}^{\mathrm{PK1}}$ does not necessarily have this property.

\section{Implementation of Constraints for Microdeformation}\label{sec:microdeformation_implementation}

Writing the isoparametric discretization 
\begin{align}
  \vec{u}=&\algmatrix{{\hat{u}}}_{(e)}\cdot\algvec{N}(\eta)\ ,&
  \vec{x}=&\algmatrix{{\hat{x}}}_{(e)}\cdot\algvec{N}(\eta)\ ,
\end{align}
in terms of a shape function vector $\algvec{N}(\eta)$ and matrices $\algmatrix{{\hat{u}}}_{(e)}$ and $\algmatrix{{\hat{x}}}_{(e)}$ of nodal displacements and nodal locations of each element $e$, respectively, the integral to compute the second geometric moment \eqref{eq:G-fullmicromorph} can be written as
\begin{align}
\tens{G}^{\chi}&
=\sum\limits_{e=1}^{n^\parallel}\algmatrix{{\hat{x}}}_{(e)}\cdot\algmatrix{{L}}_{(e)}\cdot\algmatrix{{\hat{x}}}_{(e)}^{\mathrm{T}} \ .
  \label{eq:G-fullmicromorph-FE}
\end{align}
Therein, $n^\parallel$ refers to the number of elements being located at the surface $S^\parallel$ of heterogeneity, i.e.\ at the pore surface. Furthermore, the matrix
\begin{align}
    \algmatrix{{L}}_{(e)}=\int_{S^\parallel_{(e)}}\algvec{N}\otimes\algvec{N}\,\mathrm{d}S
\end{align}
compiles the integral over the dyadic square of the shape functions along the elemental part ${S^\parallel_{(e)}}$ of the surface $S^\parallel$ of heterogeneity.
Having computed $\tens{G}^{\chi}$ in the pre-processing step, the discretized version of the micro-macro relation~\eqref{eq:voidBC-FM} for the microdeformation becomes
\begin{align}
\tens{\chi}
&=\algmatrix{{\hat{u}}}\cdot\left[ \sum\limits_{e=1}^{n^\parallel} 
\algmatrix{{A}}_{(e)}^{\mathrm{T}}\cdot\algmatrix{{L}}_{(e)}\cdot\algmatrix{{\hat{x}}}_{(e)}^{\mathrm{T}}
\right]\cdot  (\tens{G}^{\chi})^{-1}
\label{eq:voidBC-FM-FE}
\end{align}
The square bracket in \eqref{eq:voidBC-FM-FE}, together with the following second geometric moment, can be precomputed in the pre-processing step and corresponds to the coefficients of the linear constraint between microdeformation $\tens{\chi}$ and nodal displacements $\algmatrix{{\hat{u}}}$.
For this purpose, the multiplication by the Boolean node-to-element matrix $\algmatrix{A}_{(e)}$ is favorably implemented in form of an element assembly procedure as usual.

For linear elements $\algvec{N}^\mathrm{T}\!=\![1+\eta, 1-\eta]^\mathrm{T}/2$, the square matrix becomes
\begin{align}
    \algmatrix{{L}}_{(e)}&=\frac{1}{4} \begin{bmatrix} 
                                           1 & -1 \\
                                          -1 & 1 
                                        \end{bmatrix} 
                                        \Delta S
\end{align}
with element length $\Delta S=||S^\parallel_{(e)}||$.
For quadratic elements $\algvec{N}^\mathrm{T}\!=\![\eta^2-\eta, 2(1-\eta^2), \eta^2+\eta]^\mathrm{T}/2$, the result is
\begin{align}
    \algmatrix{{L}}_{(e)}&=\frac{1}{30} \begin{bmatrix} 
                                           4 & 2 & -1 \\
                                           2 & 16&  2 \\
                                           -1&  2&  4 \\
                                        \end{bmatrix} \Delta S\ ,
\end{align}
the latter being exact only if the middle node is located geometrically at the half-way position.

% References
\medskip
\bibliographystyle{unsrtnat}
\bibliography{MalikMikromorph.bib}
\end{document}